\documentclass[useAMS,usenatbib]{mnras} 
\usepackage{color}
\usepackage{times,psfig,epsfig} 
\usepackage{rotating, graphicx, adjustbox}
\usepackage{amssymb, amsmath}
\usepackage[T1]{fontenc} 
\usepackage{aecompl}
\usepackage{gensymb}

\newcommand{\alma}{{\it ALMA}}

\newcommand{\athena}{{\it ATHENA}}

\title[Cosmic magnetic fields with MASCLET]{Cosmic magnetic fields
  with MASCLET: an application to galaxy clusters} 
\author[Quilis, Mart\'{\i} \& Planelles]
 {Vicent Quilis$^{1,2}$, Jos\'e-Mar\'{\i}a Mart\'{\i}$^{1,2}$ and
   Susana Planelles$^{1}$\\ 
  $^1$Departament d'Astronomia i Astrof\'{\i}sica, Universitat de
  Val\`encia, E-46100 Burjassot (Val\`encia), Spain\\ 
  $^{2}$Observatori Astron\`omic, Universitat de Val\`encia, E-46980
  Paterna (Val\`encia), Spain} 
\date{Released \today}

\pagerange{\pageref{firstpage}--\pageref{lastpage}} \pubyear{2019}

\def\LaTeX{L\kern-.36em\raise.3ex\hbox{a}\kern-.15em
    T\kern-.1667em\lower.7ex\hbox{E}\kern-.125emX}

\begin{document}

\label{firstpage}

\maketitle

\begin{abstract}
%% Text of abstract
We describe and test a new version of the adaptive mesh refinement
(AMR) cosmological code MASCLET. The new version of the code includes
all the ingredients of its previous version plus a description of the
evolution of the magnetic field under the approximation of the ideal
magneto-hydrodynamics (MHD). To preserve the divergence-free condition
of MHD, the original divergence cleaning algorithm of \cite{DK02}
is implemented. We present a set of well-known 1D and 2D tests, such
as several shock-tube problems, the fast rotor and the Orszag-Tang vortex. 
 The performance of the code in all the tests is excellent with 
  estimated median relative errors of $\nabla \cdot {\bf B}$ in the 2D
  tests smaller than $5 \times 10^{-5}$ for the fast rotor test, and
  $5 \times 10^{-3}$ for the Orszag-Tang vortex. As an astrophysical application of the
code, we present a simulation of a cosmological box of 40  comoving
Mpc side length in which a primordial uniform comoving magnetic field
of strength 0.1 nG is seeded. The simulation shows how the magnetic
field is channelled along the filaments of gas and is concentrated and
amplified within galaxy clusters.  Comparison with the values
  expected from pure compression reveals an additional amplification
  of the magnetic field caused by turbulence in the central region of
  the cluster.  Values of the order of $\sim 1 \mu$G are obtained in clusters at $z\sim 0$  with median relative errors 
  of $\nabla \cdot {\bf B}$ below 0.4\%. The
implications of a proper description of the dynamics of the magnetic
field and their possible observational counterparts in future
facilities are discussed.
 \end{abstract}

\begin{keywords}
magneto-hydrodynamics - methods: numerical - galaxy formation
- large-scale structure of Universe - Cosmology
\end{keywords}

%% main text

\section{Introduction}
\label{s:intro}

A new era in cosmological observations favoured by forthcoming new
observational facilities, such as the {\it Square-Kilometre
  Array}\footnote{https://www.skatelescope.org/} \citep[SKA;
e.g.][]{Keshet_2004, Acosta-Pulido2015} or
\athena\footnote{http://www.the-athena-x-ray-observatory.eu}
\citep[e.g.][]{Athena_wp, Athena_2017}, and the already existing 
ones like \alma\footnote{https://www.almaobservatory.org/es/inicio/}
\citep[e.g.][]{ALMA_2009}, will produce a huge amount of extra
high-quality data, undoubtedly leading to a deeper knowledge on the
processes carving the cosmic structures in the Universe.   

The formation of galaxies and galaxy clusters would imprint features
on the baryonic component of cosmic structures that eventually  would
be   observable in different wavelengths of the electromagnetic
spectrum. Related to these processes, shock waves, turbulence,
strong gradients, and all sort of emission processes associated to the
baryonic nature of the gas are expected to be observed and quantified
by the new generation of telescopes \citep[e.g.][]{Planelles18}.  

In this context of high-quality data, the models describing the 
formation and evolution of such structures must also include all the 
relevant physical ingredients. Cosmological codes
\citep[e.g.][]{art1997, ramses2002, gadget2003, flash2005, enzo2014}
have vastly improved in the last years by including all sort of
processes associated to the microphysics of the gas \citep[see][for a
review and references therein]{Planelles_2016}. 
 Although for many
years it was believed that the role of {\bf B} was irrelevant in
the formation of galaxies and galaxy clusters, nowadays, there are
many observational and theoretical results pointing towards the
opposite direction \citep[see,
e.g.][]{Beck_2009,Govoni_2006,Bonafede_2010,Dolag_2011}. In this line,
the description of shock waves, strong gradients and X-ray emission,
among others, would require a proper modeling of the magnetic field
within current cosmological codes  to compare with the observations
delivered by the new generation of telescopes. This
  modelling should allow for the amplification of primordial weak
  magnetic fields via small-scale dynamo during the structure
  formation, and includes other astrophysical sources of magnetic
  seeding at lower redshifts (see \cite{Brandenburg_2005} 
  for a critical discussion on the origin of magnetic fields in
  clusters and a review on small-scale turbulent dynamo).

With this motivation, in recent years the most popular cosmological
codes have developed  updated versions incorporating the evolution of
seed magnetic fields \citep[see][for a recent review on cosmological
simulations including magnetic fields]{Donnert_2018}. Current
cosmological smoothed particle MHD codes include those developed by 
\citet[][the MHD extension of GADGET]{SD13} and
\citet[][GCMHD++]{BO18}. \citet{PB11} have developed the MHD extension
of AREPO (\cite{arepo2010}, a finite volume code using unstructured
moving mesh). These codes have adapted the {\it divergence cleaning}
(DC) algorithm of \cite{DK02} to remove the numerical errors which
otherwise would accumulate in non-soleinodal magnetic fields. Dedner's
DC method was also the algorithm implemented in the MHD version  of the
AMR, finite volume code ENZO \citep{WA09}. An alternative
to the DC algorithm is the {\it constrained transport} (CT) method
\citep{EH88,BS99}, which satisfies the magnetic divergence constraint
by construction. The CT method has been implemented in the AMR,
finite-volume, MHD upgrades of RAMSES \citep{FH06}, ENZO \citep{CX10}
and CHARM \citep{MM11}. AREPO-DG \citep{GP18} implements a
discontinuous Galerkin method for ideal MHD on an Eulerian AMR grid
using the {\it eight-wave method} of \cite{PR99}, an alternative DC
method to control the magnetic divergence errors.

This paper presents a new version of an already well-tested and used
cosmological code, MASCLET \citep{Quilis04}. This new version is
improved by adding the evolution of a magnetic field {\bf B} in the
context of ideal magneto-hydrodynamics (MHD) in a full cosmological
frame. Contrary to CT techniques, which evolve face-centered
  magnetic fields, the chosen DC method of \cite{DK02} keeps the
  cell-centered discretization of the full scheme (at the cost of
  introduzing an additional variable and the corresponding evolution
  equation), which makes it technically simpler to implement in the
  structure of the AMR code.

The paper is organized as follows. In  Section \ref{s:equations} the
basic equations to be solved are described. The implementation of  
the numerical techniques to solve previous equations and the details
of the numerical code are presented in Section \ref{s:numeric}. In
Section \ref{s:tests}, the code is tested by carrying out a battery of
numerical tests. A cosmological application is shown in Section
\ref{s:cluster}, where a cosmological box including a magnetic seed is
evolved. Finally, we discuss our results and offer our conclusions in
Section \ref{s:conclusion}. 

%%%%%%%%%%%%%%%%%%%%%%%%%%%
\section{Equations}
\label{s:equations}
%%%%%%%%%%%%%%%%%%%%%%%%%%%%

\subsection{The MHD equations in an expanding background}

The equations of MHD \citep{Goedbloed_2004} describing the evolution
of a magnetized fluid in a gravitational field in Eulerian coordinates
$(t, {\bf r})$ are the following: 

\begin{itemize}
\item Continuity equation:
\begin{equation}
\label{eq:continuity}
\frac{\partial \rho}{\partial t} + \nabla \cdot (\rho {\bf u}) = 0,
%\quad \mbox{\rm (continuity equation)},
\end{equation}
\item Momentum conservation:
\begin{equation}
\frac{\partial \rho {\bf u}}{\partial t} + \nabla \cdot (\rho {\bf
  u}{\bf u} + p^* {\bf I} - {\bf B}{\bf B}) = -\rho \nabla \Phi,
%\quad \mbox{\rm (momentum conservation)},
\end{equation}
\item Energy conservation:
\begin{equation}
\frac{\partial e}{\partial t} + \nabla \cdot ((e + p^*) {\bf
  u} - ({\bf B}\cdot{\bf u}){\bf B}) = -\rho {\bf v} \cdot \nabla \Phi,
%\quad \mbox{\rm (energy conservation)},
\end{equation}
\item Induction equation:
\begin{equation}
\label{eq:induction}
\frac{\partial {\bf B}}{\partial t} - \nabla \times ({\bf u} \times
{\bf B}) = 0,
%\quad \mbox{\rm (induction equation)},
\end{equation}
\item Divergence-free condition:
\begin{equation}
\label{eq:div_free}
\nabla \cdot {\bf B} = 0.
% \quad \mbox{\rm (divergence-free condition)}.
\end{equation}
\end{itemize}

In these equations, $\rho$ is the mass density, ${\bf u}$ is
the fluid velocity, ${\bf B}$ is the magnetic field, and quantities $e$
and $p^*$ stand, respectively, for the total energy density (internal
+ kinetic + magnetic) and total  pressure (thermal + magnetic):
\begin{equation}
e = \rho \epsilon + \frac{1}{2} \rho {\bf u}^2 + {\bf B}^2,
\end{equation}
\begin{equation}
p^* = p + \frac{{\bf B}^2}{2}.
\end{equation}
The thermal pressure $p$ is related to the mass density $\rho$  and
the specific internal energy $\epsilon$ by means of an equation of
state of the form $p = p(\rho, \epsilon)$. Quantity $\Phi$ appearing
in he source terms of the momentum and energy equations stands for the 
gravitational potential.

In the presence of an expanding background, the system of
Eqs.~\ref{eq:continuity}-\ref{eq:div_free} is rewritten in terms of
the comoving coordinates, ${\bf x} \equiv {\bf r}/a(t)$, where $a(t)$
is the scale factor and $H(t) \equiv \dot{a}/{a}$ is the Hubble
constant \citep{Pe80}: 

\begin{equation}
\label{eq:continuity_q}
\frac{\partial \tilde{\rho}}{\partial t} + \frac{1}{a}\nabla \cdot
(\tilde{\rho} {\bf 
  v}) = 0,
\end{equation}
\begin{equation}
\label{eq:momentum_q}
\frac{\partial \tilde{\rho} {\bf v}}{\partial t} + \frac{1}{a} \nabla
\cdot (\tilde{\rho} {\bf
  v}{\bf v} + \tilde{p}^* {\bf I} - \tilde{\bf B}\tilde{\bf B}) = -
H\tilde{\rho} {\bf v} - \frac{\tilde\rho}{a}\nabla\phi 
\end{equation}
\begin{eqnarray}
\label{eq:energy_q}
\nonumber
\frac{\partial \tilde{E}}{\partial t} \!\!\!& \!\!+\!\!& \!\!\!
\frac{1}{a} \nabla \cdot ((\tilde{E} + \tilde{p}^*) {\bf 
  v} - (\tilde{\bf B}\cdot{\bf v})\tilde{\bf B}) = \\
& & \ \ \ \ \ \ = - 3H \tilde{p}^*
+ H \tilde{\bf B}^2 - H \tilde{\rho} {\bf v}^2 -
\frac{\tilde\rho}{a} {\bf v} \cdot \nabla\phi, 
\end{eqnarray}
\begin{equation}
\label{eq:induction_tilde}
\frac{\partial \tilde{\bf B}}{\partial t} - \frac{1}{a} \nabla \times
({\bf v} \times 
\tilde{\bf B}) = - \frac{H}{2} \tilde{\bf B}, 
\end{equation}
\begin{equation}
\label{eq:div_free_tilde}
\nabla \cdot \tilde{\bf B} = 0.
\end{equation}
In this set of equations, all the spatial differential operators
\textcolor{blue} {refer to} the comoving coordinates ${\bf x}$. 
The overdensity with respect to the background density, $\rho_{_{B}}$,
is  $\tilde\rho = \rho/\rho_{{B}}$. The peculiar velocity of the
fluid  is  ${\bf v} = {\bf u} -\dot a {\bf  x}$, being $\dot a {\bf
  x}$  the Hubble flow velocity. The quantity $\tilde{
E}=E/\rho_{B}$ is related to the total energy density, $E$,
of the magneto-fluid, which includes only the kinetic energy
corresponding to the fluid's peculiar velocity, i.e., $E = \rho
\epsilon + \frac{1}{2} \rho {\bf v}^2 + {\bf B}^2$. The total pressure
and the magnetic field are  redefined as $\tilde{p}^*= p^*/\rho_{B}$ 
 and $\tilde{\bf B} = {\bf B}/\sqrt{\rho_{\rm B}}$, respectively. 
 
The peculiar potential, $\phi = \Phi + a \"a {\bf x}^2/2$, satisfies
the Poisson equation 
\begin{equation}
{\Delta \phi} = \frac{3}{2}H^2a^2 \Omega \tilde \rho
\end{equation}
where $\Omega$ is the density parameter. 

Equations~(\ref{eq:continuity_q}-\ref{eq:energy_q}) coincide with
those of \cite{Quilis04} in the case with $\tilde{\bf B} = 0$.

\subsection{Divergence cleaning}

MASCLET uses the divergence cleaning algorithm of \cite{DK02} to
control the magnetic field divergence errors. In the original mixed
hyperbolic/parabolic correction approach of \cite{DK02} (the one
chosen in MASCLET), the divergence constraint of the magnetic field
and the induction equation (in an Eulerian frame,
Eqs.~\ref{eq:induction}-\ref{eq:div_free}) are coupled by introducing
an additional scalar function $\psi$ in such a way that: i) $\nabla
\cdot {\bf B}$ and $\psi$ satisfy the same evolution equation, and ii)
this evolution equation leads to the propagation to the domain
boundaries and the decay of $\nabla \cdot {\bf B}$ (and $\psi$).

In the case of an expanding background, the original Dedner et al.'s
approach can be applied once the linear $\tilde{\bf B}$-term in
Eq.~\ref{eq:induction_tilde} is removed by shifting to the comoving
magnetic field ${\bf B}' = a^2 \sqrt{\rho_{\rm B}} \ \tilde{\bf B}$,
and the $\psi$-terms are introduced in
Eqs.~(\ref{eq:induction_tilde}-\ref{eq:div_free_tilde})  
\begin{equation}
\label{eq:induction_dedner}
\frac{\partial {\bf B}'}{\partial t} - \frac{1}{a} \nabla \times ({\bf
  v} \times {\bf B}') + \nabla \psi = 0,
\end{equation}
\begin{equation}
\label{eq:div_free_dedner}
\frac{\partial \psi}{\partial t}  + c_h^2 \nabla \cdot {\bf B}' = -
\frac{c_h^2}{c_p^2} 
\psi.
\end{equation}

Now, direct manipulation of these two equations leads to the {\it
telegraph equation} for $\psi$ (and $\nabla \cdot {\bf B}'$)
\begin{equation}
\label{eq:telegraph}
\frac{\partial^2 \psi}{\partial t^2} + \frac{c_h^2}{c_p^2}
\frac{\partial \psi}{\partial t} = c_h^2 \Delta \psi
\end{equation}
establishing that divergence errors propagate away from the point
where they are produced at a speed $c_h$, and damp at a rate given by
$c_h^2/c_p^2$. Quantities $c_h$ and $c_p$ are dimensional parameters
to be tuned (see next section for details).

Equations (\ref{eq:induction_dedner}) and (\ref{eq:div_free_dedner})
substitute the original ones (Eqs.~\ref{eq:induction_tilde} and
\ref{eq:div_free_tilde}) in the system of equations. Since they are
designed to ensure that $\nabla \cdot {\bf B}'$ (as $\psi$) verify the
telegraph equation, the errors in $\nabla \cdot {\bf B}'$ propagate
out of the numerical domain and damp as desired.  Since $\displaystyle
\nabla \cdot \tilde{{\bf B}} = \frac{1}{a^2 \sqrt{\rho_{\rm B}}}
\nabla \cdot {\bf B}' \propto (\nabla \cdot {\bf B}')/\sqrt{a}$,
errors in $\nabla \cdot \tilde{{\bf B}}$ dilute in the expanding
background as they propagate and damp.

%%%%%%%%%%%%%%%%%%%%%%%%%
\section{Numerical implementation}
\label{s:numeric}
%%%%%%%%%%%%%%%%%%%%%%%%%

\subsection{The reference version of  MASCLET code}

MASCLET \citep[Mesh Adaptive Scheme for CosmologicaL 
  structurE evoluTion;][]{Quilis04}, is a cosmological
  multidimensional hydrodynamic and N-body code based on an adaptive 
  mesh refinement scheme (AMR). We address interested readers to
    the original paper for a complete technical
    description of the code and offer here a short summary of its
    basic ingredients. 

The hydro solver is a high-order, finite volume Godunov scheme based on
monotonicity preserving cell-reconstruction routines and approximate
Riemann solvers. Two different cell reconstructions are implemented:
a piecewise linear reconstruction with the MINMOD slope limiter, and
the piecewise parabolic method PPM. Two Riemann solvers are also
implemented: a Roe-type linearized Riemann solver and the HLLE Riemann 
solver. Advance in time is done by TVD preserving second and third
order Runke-Kutta methods following a method of lines.

Dark matter is evolved using a Particle-Mesh scheme with a
Lax-Wendroff temporal integrator. 

Both components, gas and dark matter, evolve coupled by the total
gravitational field which is computed at each time step by solving the
Poisson equation. 

The gain in numerical resolution, both spatial and temporal, is
obtained by means of an AMR scheme. This method
refines the original coarse grid into patches whose cells are half size
of their parent cells. In these new grids, all the relevant physical
quantities are be obtained, either by evolution of the same quantities
from the previous time step, or by tri-linear interpolation from the parent
grid (lower level). This process can be repeated iteratively between
two consecutive levels, producing a whole hierarchy of nested
grids. The criteria to decide which regions of a 
given grid must be refined,  can be configured for every specific
application \citep[see, e.g.][]{Ricciardelli_2013,Quilis_2017,Planelles18}.  

Once the AMR hierarchy is defined, at each level, the hydro and the
dark matter solvers, previously described, can be applied. The
different patches at the same level and the child patches and their
parents are connected through the boundary conditions and the values
used to initialize them.  

In the same manner, a multigrid SOR method is used to solve the
Poisson equation in each patch of the hierarchy.    

The current version of the code also  includes  inverse Compton and
free-free cooling, UV heating, atomic and molecular cooling for the
gas depending on the metallicity,  and a phenomenological description of
star formation. The description of feedback phenomena, both thermal
and kinematic, from stars and active galactic nuclei is also included.  

\subsection{The MHD version of  MASCLET code}

The MHD version of MASCLET solves the system formed by
Eqs.~(\ref{eq:continuity_q}-\ref{eq:energy_q}),
(\ref{eq:induction_dedner}) and (\ref{eq:div_free_dedner})
incorporating all the numerical ingredients of the reference code with
the only evident modifications of the fluxes in
Eqs.~(\ref{eq:continuity_q})-(\ref{eq:energy_q}), and the addition of
two new equations
(Eqs.~\ref{eq:induction_dedner}-\ref{eq:div_free_dedner}).   

These equations are evolved in conservation form with
numerical fluxes which, in general, depend on the eigenstructure of the
system. Following the discussion in \cite{DK02}, in the
one-dimensional case (the one used to obtain the eigenstructure),
the system decouples into a subsystem formed by the equations for
the parallel component of the magnetic field, in our case
${{B}'_\parallel}$, and ${\psi}$, and the usual 1D-MHD system (with a
factor $1/a$ in front of the vector of fluxes) for the variables
$(\tilde{\rho}, {\bf v}, \tilde{\bf B}_\perp, \tilde{p})$ with the
magnetic field divergence-free constraint, $\tilde{B}_\parallel$
constant.  

  Along the $x$-direction, for sufficiently large $c_h$, the
corresponding eigenvalues of the full system, ordered in a
nondecreasing sequence, are:  
$$
\lambda_1 = -c_h, \  \lambda_2 = (v^x - {v}_f)/a, \ \lambda_3 = (v^x
  - {v}_a)/a, 
$$
$$
\lambda_4 = (v^x - {v}_s)/a, \ \lambda_5 =
      v^x/a, \ \lambda_6 = (v^x + {v}_s)/a, 
$$
$$
\lambda_7 = (v^x
  + {v}_a)/a, \ \lambda_8 = (v^x + {v}_f)/a, \ \lambda_9 = c_h\, ,
$$
where
\begin{equation}
{v}_a = \frac{\tilde{B}^x}{\sqrt{\tilde{\rho}}}
\end{equation}
and
\begin{equation}
v_{f,s}
= \left\{ \frac{1}{2}\left( {c}_s^2 + \frac{\tilde{\bf
          B}^2}{\tilde{\rho}} \pm \sqrt{\left( {c}_s^2 + \frac{\tilde{\bf
          B}^2}{\tilde{\rho}}\right)^2 - 4 {c}_s^2 {v}_a^2} \right)\right\}^{1/2}
\end{equation}
are, respectively, the Alfv\'en speed and the fast and slow
magnetosonic speeds, and ${c}_s$ is the sound speed.

On the other hand, for any right eigenvector $\tilde{\bf r}_i =
(\tilde{r}^1_{\,i},..., \tilde{r}^7_{\,i})^T$ of the original  1D-MHD system  
\citep[Eqs.~\ref{eq:continuity_q}-\ref{eq:energy_q} and components $y$
and $z$ of Eq.~\ref{eq:induction_dedner}; see e.g.][]{BW88}, there is
a right eigenvector of the full MHD system with the divergence
correction (Eqs.~\ref{eq:continuity_q}-\ref{eq:energy_q},
\ref{eq:induction_dedner} and \ref{eq:div_free_dedner})
$\tilde{\bf R}_i = (\tilde{r}^1_{\,i},..., \tilde{r}^5_{\,i}, 0,
\tilde{r}^6_{\,i}, \tilde{r}^7_{\,i}, 0 )^T$ with respect to the same
eigenvalue $\lambda_i$, $i = 2,3,...,8$. Moreover, two new
eigenvectors appear associated to  eigenvalues $\lambda _1$ and
$\lambda_9$, respectively, $\tilde{\bf R}_1 = (0,0,0,0,0,1,0,0,-c_h)$
and $\tilde{\bf R}_9 = (0,0,0,0,0,1,0,0,c_h)$.

In its present version, MASCLET incorporates a new slope limiter 
\citep[MC, monotoniced central-difference limiter; see for
instance][]{Mignone_2006} for the piecewise linear reconstruction, 
and an HLLE Riemann solver adapted to the MHD equations. All the
results shown in this paper have been obtained with the MC
slope limiter. The 1D subsystem for the variables
$(\tilde{\rho}, {\bf v}, \tilde{\bf B}_\perp, \tilde{p})$ is evolved
in time with numerical fluxes calculated with the HLLE Riemann solver 
based on upper and lower bounds, respectively, of the largest and
smallest local propagation speed of fast magnetosonic waves,
$\lambda_8$, $\lambda_2$. The decoupled subsystem for
${{B}'_\parallel}$ and ${\psi}$ (in charge of the diverence cleaning)
is then evolved in time with numerical fluxes based on a (constant)
characteristic speed $c_h$ chosen to be the maximum speed compatible
with the time step restriction, i.e., fixed to the largest (in
absolute value) propagation speed of fast magnetosonic waves,
$\lambda_2$, $\lambda_8$, across the grid. Following \cite{MT10}, and
attending to the dimensional nature of $c_h$ and $c_p$, in several 
preliminary tests we defined $c_p = \sqrt{\Delta x c_h/\alpha}$, with
$\alpha = 0.2, 0.5$ and $\Delta x$ being the cell size of the finest
grid. The results in the AMR applications were not satisfactory and
numerical instabilities between different levels of resolution arose.
The final choice was to recover Dedner's original prescription
\citep{DK02} as suggested recently by \cite{GP18} in the context on
non-uniform grids, and take $c_p = \sqrt{0.18 c_h}$  (in code units). This choice of
$c_h$ and $c_p$  is the one used in all the numerical applications
presented in this paper.

%%%%%%%%%%%%%%%%%%%%%%%%%%
\section{Tests}
\label{s:tests}
%%%%%%%%%%%%%%%%%%%%%%%%%

In this Section, we present several classical MHD tests in order to
quantify the performance of the code. In all of them, there is no
expanding background as in the general form of
Eqs.~\ref{eq:continuity_q}-\ref{eq:div_free_tilde}. Consequently,
these tests have been carried out by integrating the aforementioned
equations with $a=1$, $H=0$ and $\rho_B = 1$ (in the absence of any 
gravitational field).  

\subsection{Shock-tube tests}
\label{ss:tube}

Shock tubes have become standard tests where to prove the ability of 
hydrodynamical codes to describe shocks and contact
discontinuities. In the case of MHD codes, the importance of shock
tube tests is even greater as the solution involves a larger variety
of discontinuity types \citep[see, e.g.][]{JT64, TY14}. In this section we
discuss the performance of  MASCLET in reproducing  two
representative MHD shock tube tests from \cite{RJ95}, in particular, 
test 2a and test 4a. In both tests, we set the adiabatic index of the 
equation of state to $\gamma = 5/3$, and use a one-dimensional grid
with $x \in [0,1]$. The initial discontinuity is placed at $x=0.5$. 
  
  The tests are performed on a fixed grid with a numerical resolution of
512 zones and a piecewise linear reconstruction with the MC slope limiter. 
Finally, in one-dimensional tests, the divergence free
constraint is fulfilled automatically and the divergence cleaning
algorithm does not operate. The numerical results are plotted as red
circles whereas the analytical solutions (kindly provided by D. Ryu)
appear as filled lines.

\begin{figure*}
\centering
\includegraphics[width=13.cm,angle=90]{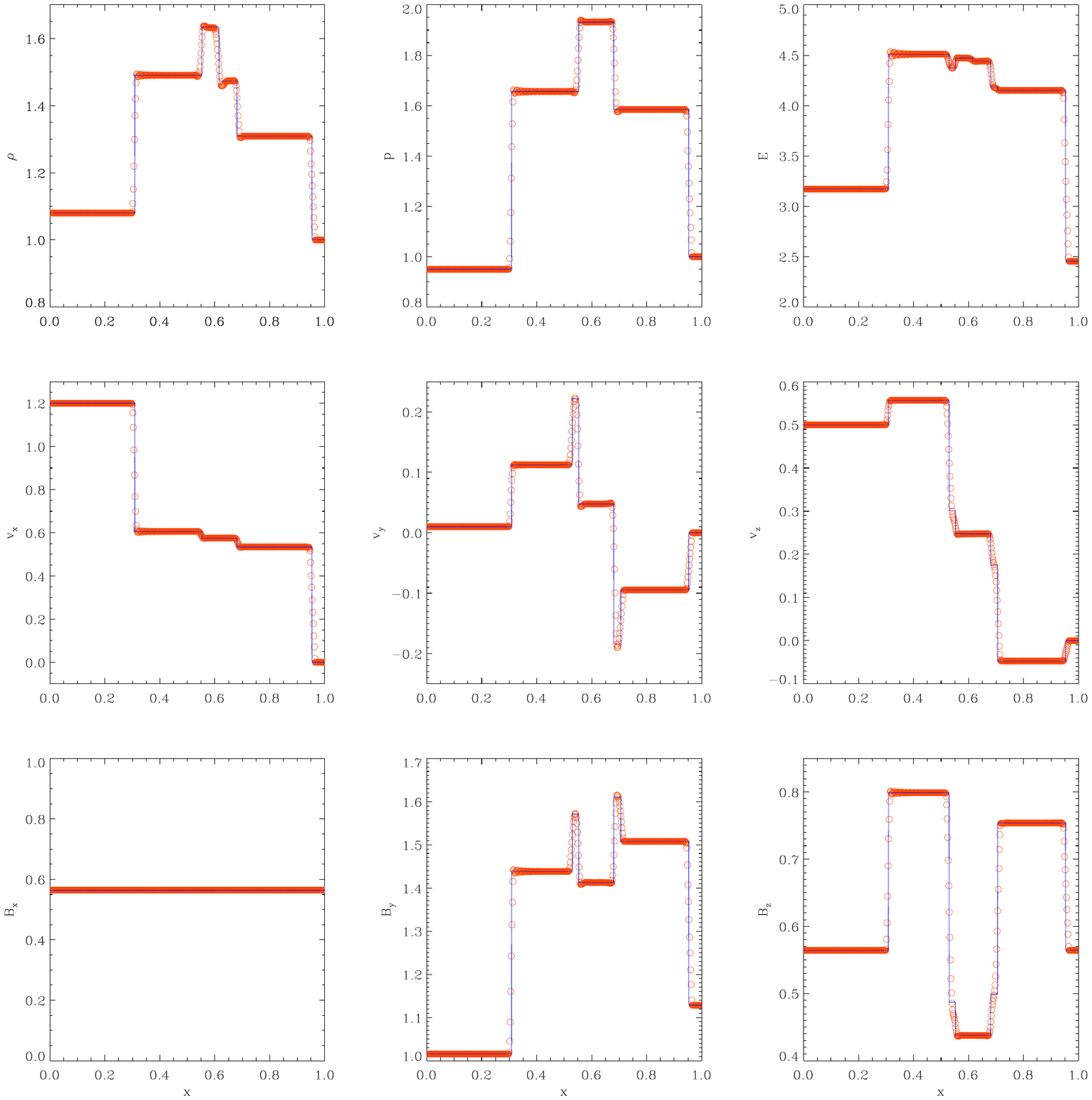}
\caption{Test 2a of \citet{RJ95} (initial data: see Section \ref{ss:tube}). Fast shocks,
rotational discontinuities and slow shocks propagate from each side of
the contact discontinuity. From top to bottom and from left to 
  right, the panels show the mass density, gas pressure, total energy,
the three components ($x$, $y$ and $z$) of the flow velocity and the
three components of the magnetic field at $t = 0.2$. Numerical results
(red circles) are plotted on top of the analytical solution (blue line).}
\label{f1}
\end{figure*}

The first test is test 2a in \cite{RJ95} with left state $\{ \rho_L,
v^x_L, v^y_L, v^z_L, B^y_L, B^z_L, p_L\} = \{ 1.08, 1.2, 0.01, 0.5, 3.6/\sqrt{4
  \pi}, 2/\sqrt{4 \pi}, 0.95\}$ and right state $\{ \rho_R,
v^x_R, v^y_R, v^z_R, B^y_R, B^z_R, p_R\} = \{ 1, 0, 0, 0, 4/\sqrt{4
  \pi}, 2/\sqrt{4 \pi}, 1\}$, and $B^x = 2/\sqrt{4 \pi}$, involving a
three-dimensional magnetic field and flow velocity. The solution at $t
= 0.2$ is shown in Fig.~\ref{f1}. Fast shocks, rotational
discontinuities (where the transverse magnetic field changes its
direction) and slow shocks propagate from each side of the contact
discontinuity. Despite the thinness of some structures, all the
constant states and discontinuities are properly captured   and only
tiny numerical oscillations and overshootings are seen in some
quantities in the post-shock states. 

%%%%%%%%%%%%%%%%%%%%%%%%%%%%%%%%%%%%%%%%%%%%%%%%%%%%%%%%%%%
%
%
%%%%%%%%%%%%%%%%%%%%%%%%%%%%%%%%%%%%%%%%%%%%%%%%%%%%%%%%%%%

\begin{figure*}
\centering
\includegraphics[width=12.cm,angle=90]{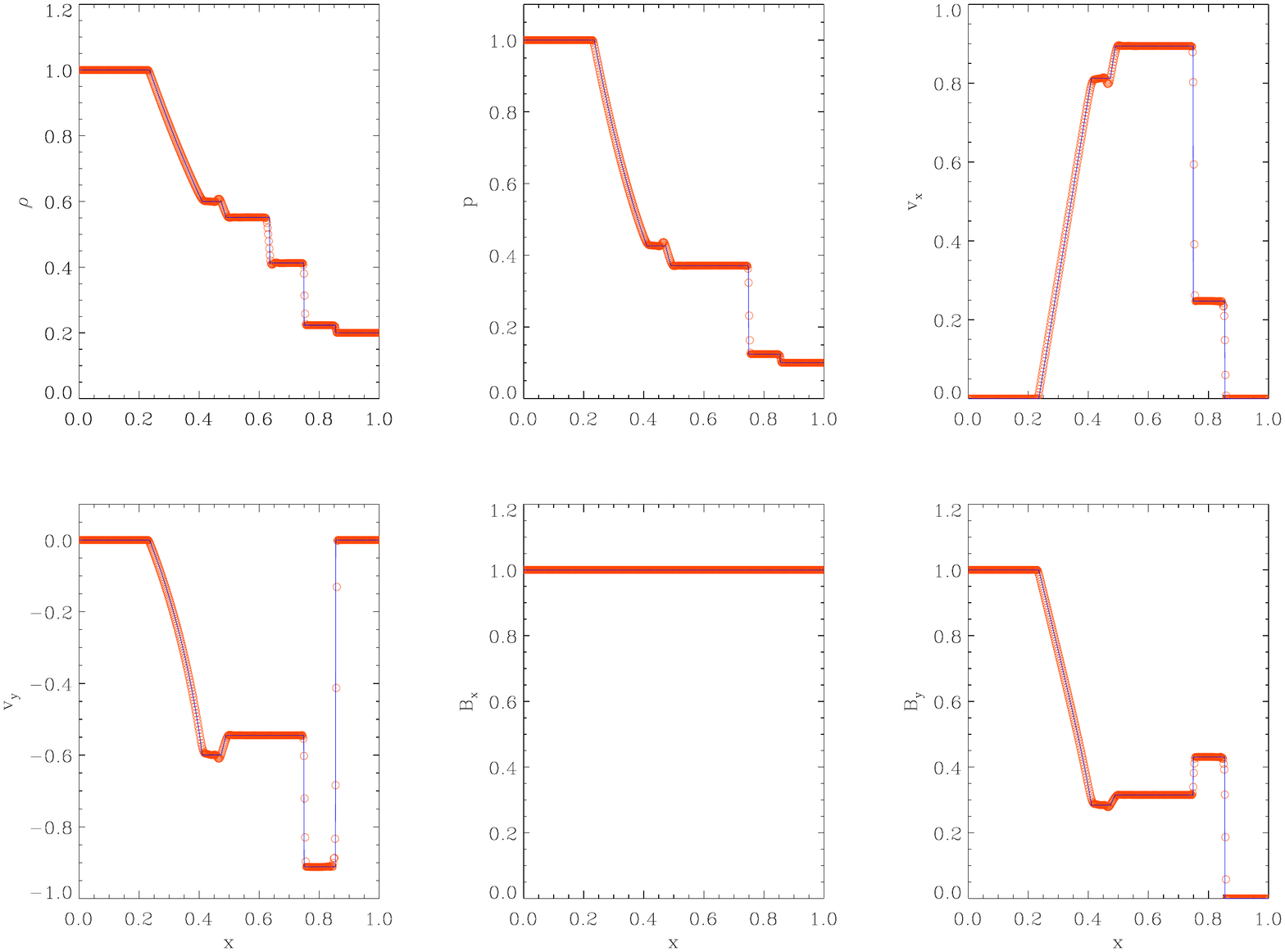}
\caption{Test 4a of \citet{RJ95} (planar Riemann problem; initial data:
  see Section \ref{ss:tube}). In this test, the initial transverse magnetic fields and
  velocities are confined in a plane. The solution produces
  left-propagating fast and slow rarefactions, a contact
  discontinuity, and right-propagating slow and fast (switch-on)
  shocks. From top to bottom and from left to right, the panels show
  the mass density, gas pressure, and $x$ and $y$ components of the
  flow velocity and the magnetic field at $t = 0.15$. Numerical
  results (red circles) are plotted on top of the analytical solution
  (blue line).} 
\label{f2}
\end{figure*}

The second test is test 4a in \cite{RJ95} with left state $\{ \rho_L,
v^x_L, v^y_L, v^z_L, B^y_L, B^z_L, p_L\} = \{ 1, 0, 0, 0, 1, 0,
1\}$ and right state $\{ \rho_R, v^x_R, v^y_R, v^z_R, B^y_R, B^z_R,
p_R\} = \{ 0.2, 0, 0, 0, 0, 0, 0.1\}$, and $B^x = 1$. The solution at
$t = 0.15$ is shown in Fig.~\ref{f2}. This test is a {\it planar} Riemann
problem in which the initial transverse magnetic fields and velocities
are confined in a plane. In this case, the Alfv\'en waves, which
rotate the fields, do not emerge. From left to right, the solution
produces a fast and a slow rarefaction, a contact discontinuity, a
slow shock and a fast, {\it switch-on} shock (where the tranverse
magnetic field turns on behind the shock). As in the previous case,
the code captures the analytical solution correctly with only small
overshootings at discontinuities and at the end points of the
rarefaction waves.

%%%%%%%%%%%%%%%%%%%%%%%%%%%%%%%%%%%%%%%%%%%%%%%%%%%%%%%%%%%
%
%
%%%%%%%%%%%%%%%%%%%%%%%%%%%%%%%%%%%%%%%%%%%%%%%%%%%%%%%%%%%

\subsection{The fast rotor}
\label{ss:rotor}

The rotor problem \citep{BS99} aims to test the propagation of strong
torsional Alfv\'en waves. A dense, rapidly spinning disk rotates
rigidly in a light static ambient medium. Both disk and ambient fluids are
threaded by an initially uniform magnetic field. The rapid rotation of
the rotor causes torsional Alfv\'en waves to propagate into the
ambient fluid reducing the angular momentum of the rotor and hence its 
rotation velocity. At the same time, the magnetic field lines wrapped
along the surface of the rotor increase the magnetic pressure
compressing the fluid in the rotor and giving it an oblong shape.

The numerical domain for this test consists of a two-dimensional
squared grid of unit length with zero-gradient boundary conditions. A
rotating disk of radius $r_0$ is embedded in a homogeneous ambient
fluid with a transition layer between $r_0$ and $r_1$ ($> r_0$). In
Cartesian $(x,y)$ coordinates, the initial conditions are given by
$\{\rho, v^x, v^y\} = \{10, -\omega \, y, \omega \, 
x\}$ for $r = \sqrt{x^2+y^2} \leq r_0$, and $\{\rho, v^x, v^y\} =
\{1,0,0\}$ for $r \geq r_1$. For $r \in ]r_0, r_1[$ a transition layer
is set with $\{\rho, v^x, v^y\} = \{1+9f, -f \; \omega y r_0/r, f
\omega x r_0/r\}$ where $\omega = 20$, $f = (r_1 - r)/(r_1 - r_0)$,
and $r_0 = 0.1$, $r_1 = 0.115$. The thermal pressure is constant with
$p = 1$ throughout the grid and the magnetic field is uniform and
aligned with the $x$ axis ($B^x = 5/\sqrt{4 \pi}$). An ideal gas
equation of state with $\gamma = 1.4$ is used. As in the previous
tests, $a$, $H$ and $\rho_B$ are set to 1, 0, 1, respectively.

\begin{figure}
\centering
\includegraphics[width= 8.5 cm,angle=0]{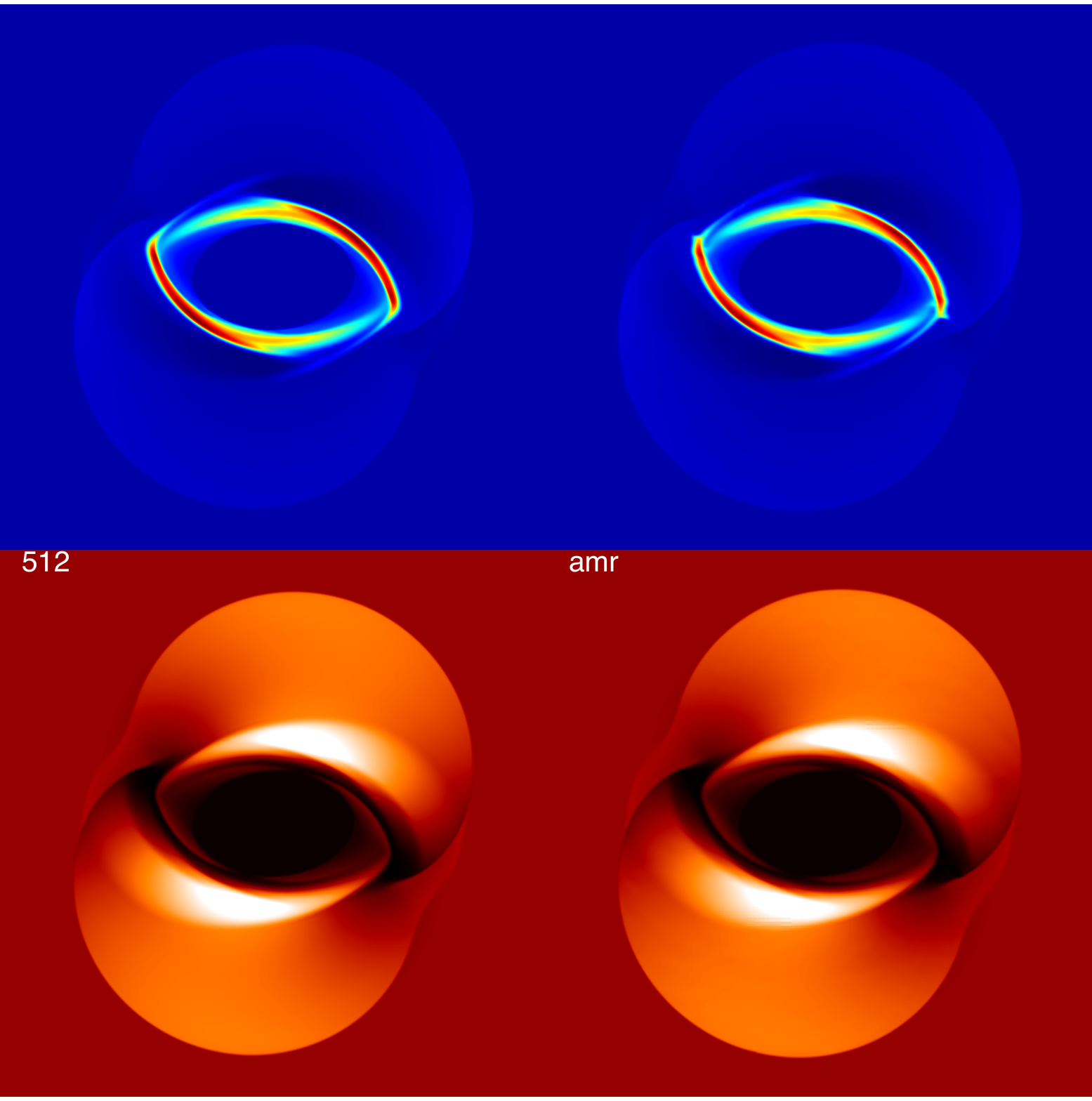} 
\caption{Rotor test. The figure shows the density (top row) and
  magnetic pressure -- in arbitrary units -- distributions (bottom
  row) at time $t = 0.15$ with two 
  numerical resolutions: a fixed grid with $512 \times 512$ cells
  (left) and an AMR grid with a coarse grid of $128 \times 128$ cells
  and two levels of refinement (refinement factor 2).}
\label{f3-panel}
\end{figure}

Figure~\ref{f3-panel} shows the density (top row) and  magnetic pressure
(bottom row) distributions at time $t = 0.15$ (when the torsional
Alfv\'en waves are about to reach the grid boundaries) with two
numerical resolutions: a fixed grid with $512 \times 512$ cells and an
AMR grid with a coarse grid of $128 \times 128$ cells and two levels of
refinement (refinement factor 2; effective resolution $512 \times 512$ 
cells). The piecewise linear reconstruction with the MC limiter is used.

Both simulations resolve the structure of the flow (high-density
shell at the rotor's oblong surface just behind the high-magnetic
pressure barrier; outgoing torsional Alfv\'en waves) with the same
detail \citep[see, e.g.][for comparison]{MT10,GP18}. 

\begin{figure}
\centering
\includegraphics[width=9.cm,angle=0]{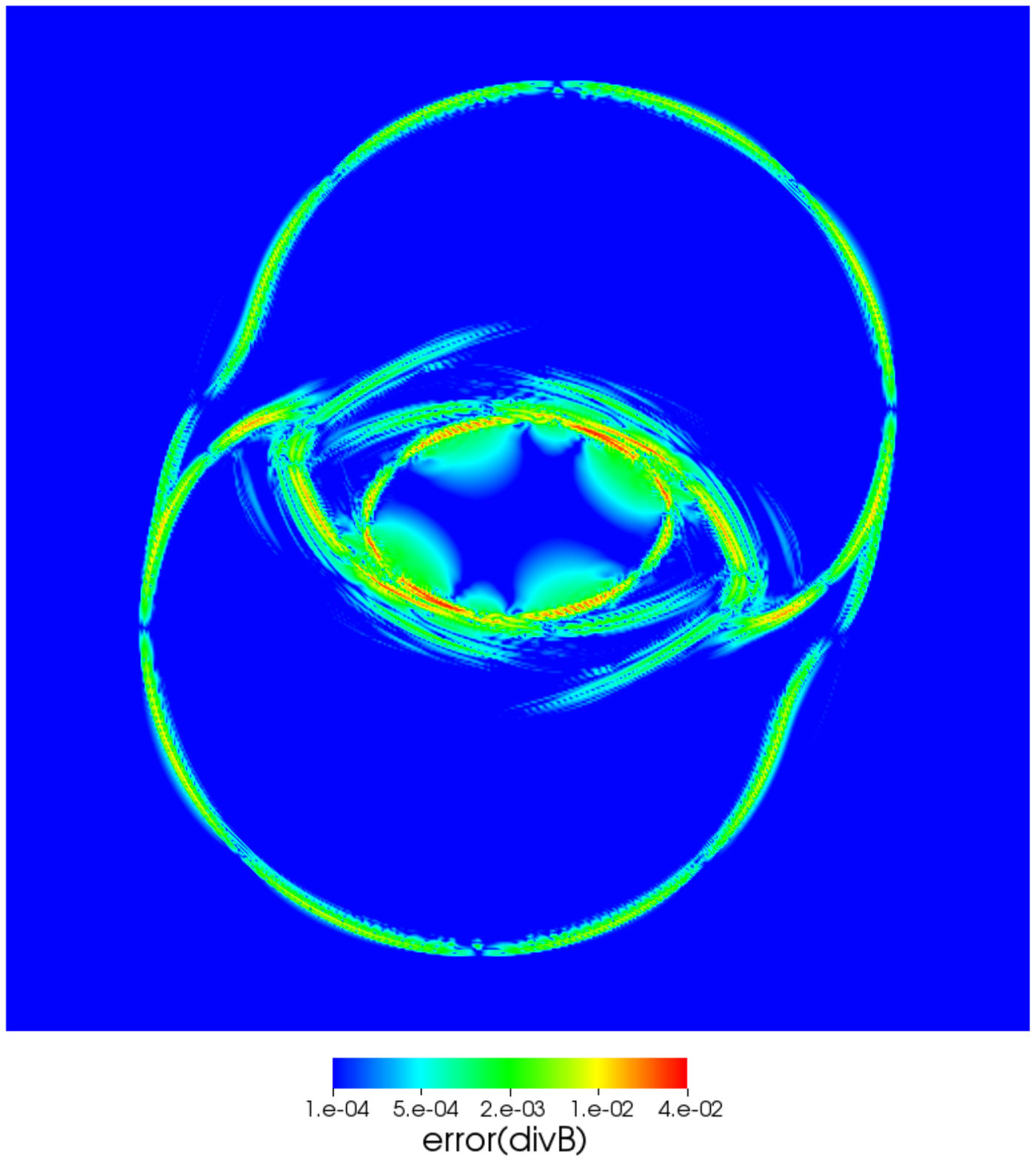} 
\caption{Map of the relative error of $\nabla \cdot {\bf B}$ (see text for the
  definition) in the rotor test at 0.15 time units for our simulations at the best resolution.}
\label{f4}
\end{figure}

Following, for instance \cite{CX10} and \cite{SD13}, we define a
  dimensionless quantity that could be interpreted as a  
relative error of the divergence of  the magnetic field, 
 $|\nabla \cdot {\bf B}_i| \Delta x_i/|{\bf B}_i|$, where
${\bf B}_i$ is the magnetic field in the $i$ cell, $\nabla \cdot {\bf
  B}_i$ is the value of the divergence computed numerically in that
cell, and $\Delta x_i$ is the cell width. In order to show the
performance of our numerical scheme, we present Figure~\ref{f4} where    
a map of the relative error of the $\nabla \cdot {\bf B}$ for our
best numerical resolution run is shown at 0.15 time units. Largest
errors appear associated to and move with magnetic field
discontinuities. 

\begin{figure}
\centering
\includegraphics[width=8.cm,angle=0]{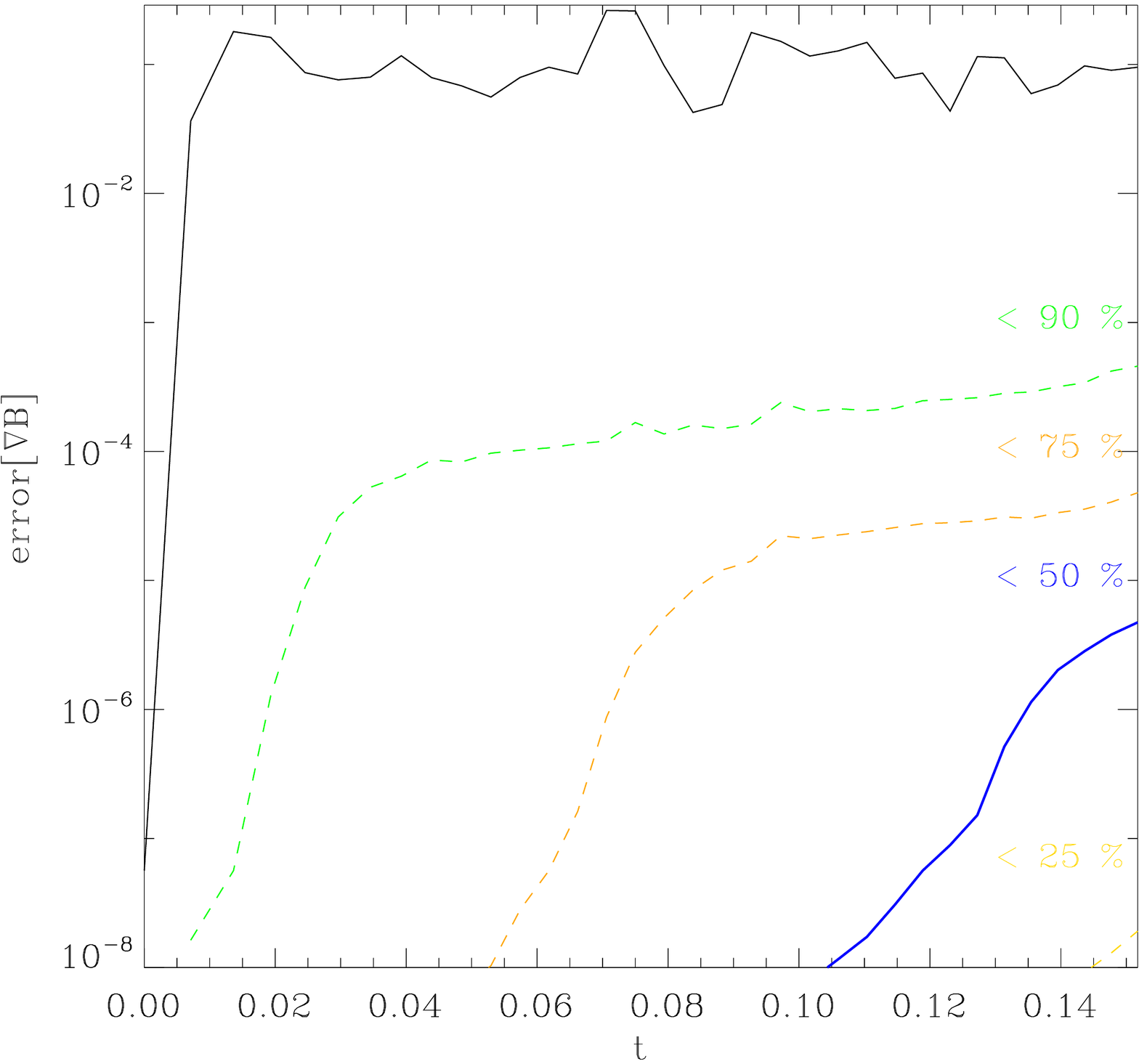} 
\caption{Relative error of $\nabla \cdot {\bf B}$ (see text for the
  definition) in the rotor test as a function of time. Yellow, 
  blue, orange and green lines indicate the upper bound error for 25,
  50, 75 and 90\% of the cells. The black line represents the maximum
  error of the divergence.}
\label{f5}
\end{figure}

Complementary to Fig.~\ref{f4} which offers a description of the
distribution and magnitude of the numerical error of the magnetic
field divergence, we present a statistical analysis in
Figure~\ref{f5}, where yellow, blue, orange and green lines indicate
the upper bound errors for 25, 50, 75 and 90\% of the cells as a
function of time, in the AMR run. Most of the cells (more than 90\%)
have divergence errors smaller than $10^{-3}$ along the
simulation. The maximum error of the divergence (black line) remains
stable at about 10\%. These errors values are in fair agreement with
those shown by \cite{GP18} for an AMR run using divergence cleaning.

%%%%%%%%%%%%%%%%%%%%%%%%%%%%%%%%%%%%%%%%%%%%%%%%%%%%%%%%%%%
%
%
%%%%%%%%%%%%%%%%%%%%%%%%%%%%%%%%%%%%%%%%%%%%%%%%%%%%%%%%%%%

%%%%%%%%%%%%%%%%%%%%%%%%%%%%%%%%%%%%%%%%%%%%%%%%%%%%%%%%%%%
%

%%%%%%%%%%%%%%%%%%%%%%%%%%%%%%%%%%%%%%%%%%%%%%%%%%%%%%%%%%%

\subsection{Orszag-Tang vortex}

The initial state of the Orszag-Tang vortex \citep{OT79} consists of
two magnetic field loops embedded into a large-scale rotating flow
structure on a two-dimensional box of unit length with periodic
boundary conditions. In the subsequent evolution, complex, small-scale
structures are formed. It has become a standard test to probe the
accuracy of MHD codes when simulating the formation and
interaction of MHD shock waves and the transition to two-dimensional MHD
turbulence. 

\begin{figure*}
\centering
\includegraphics[width=12 cm,angle=90]{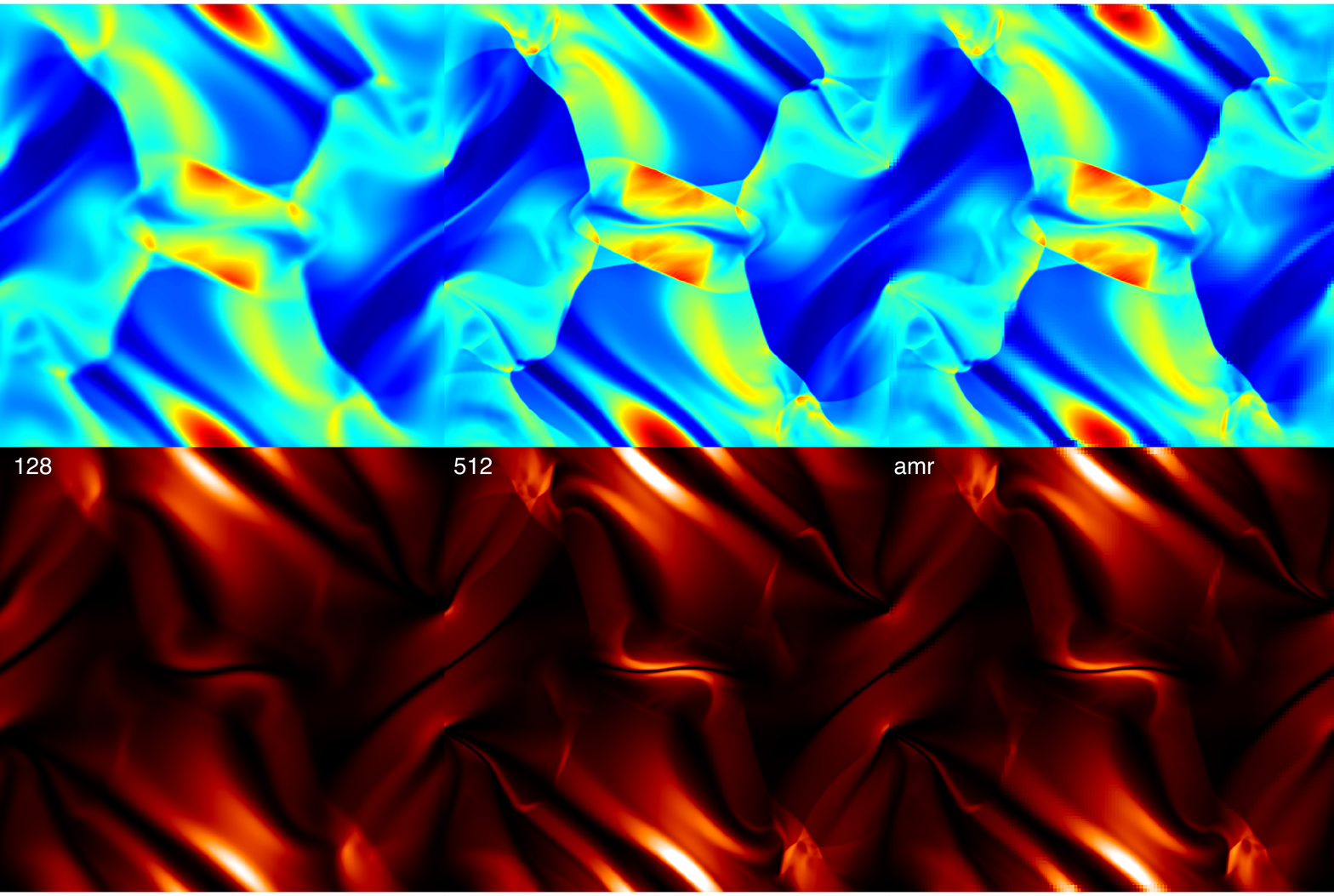} 
\caption{Orszag-Tang test. The figure shows the density (top row) and
  magnetic pressure (bottom row)  -- in arbitrary units --
  distributions at time $t = 0.5$ with three numerical
  resolutions. From left to right: a fixed grid of $128 \times 128$
  cells, a fixed grid with $512 \times 512$ cells and an AMR grid with
  a base grid of $128 \times 128$ cells and two additional levels of 
refinement with a refinement factor of 2.}
\label{f6-panel}
\end{figure*}

In Cartesian $(x,y)$ coordinates, the initial conditions are  $\{ \rho,
v^x, v^y, B^x, B^y, p\} = \{ \rho_0, -v_0 \sin (2 \pi y), v_0
\sin (2 \pi x), -B_0 \sin (2 \pi y), B_0 \sin (4 \pi x), p_0\}$
with $\rho_0 = 25/(36 \pi)$, $v_0 = 1$, $B_0= 1/\sqrt{4 \pi}$, $p_0 =
5/(12 \pi)$, and an ideal gas equation of state with adiabatic index
$\gamma = 5/3$. As in the previous tests, we set $\{a, H, \rho_B\} =
\{1,0,1\}$ in the code equations. Figure~\ref{f6-panel} shows
the density (top row) and  magnetic pressure (bottom row)
distributions at time $t = 0.5$ with three numerical resolutions: a
fixed grid of $128 \times 128$ cells, a fixed grid with $512 \times
512$ cells and an AMR grid with a coarse grid of $128 \times 128$ cells
and two additional levels of refinement with a refinement factor of 2
(for an effective resolution of $512 \times 512$ cells). The piecewise
linear reconstruction with the MC limiter is used.

As it is clearly seen, the AMR run captures the small structures with
the same degree of detail than the $512 \times 512 $ cells fixed grid
run. Additionally, the agreement between our result and previous
published works  is excellent \citep[e.g.][]{RM98,LD00,FH06,CX10,GP18}. 

{The zero-gradient boundary conditions used, as well as
  the fact that a non-negligible fraction of the 
  cells are only slightly perturbed along the simulation, favour that
  the magnetic field divergence errors remain small in the rotor test. In the
  Orszag-Tang vortex test, on the contrary, all the cells are largely
  perturbed in the course of the evolution. This, together with the
  periodic boundary conditions used in this test (similar to those used
  in standard cosmological simulations), makes the Orszag-Tang vortex
  harder from the point of view of the magnetic divergence
  errors\footnote{{Whereas the zero-gradient boundary 
    conditions used in the rotor test let the outgoing divergence
    errors driven by the hyperbolic term leave the numerical grid, the
    periodic boundary conditions imposed in the Orszag-Tang vortex
    reintroduce them continuously into the computational domain
    (although damped).}} but
  also more meaningful.}

\begin{figure}
\centering
\includegraphics[width=9.cm,angle=0]{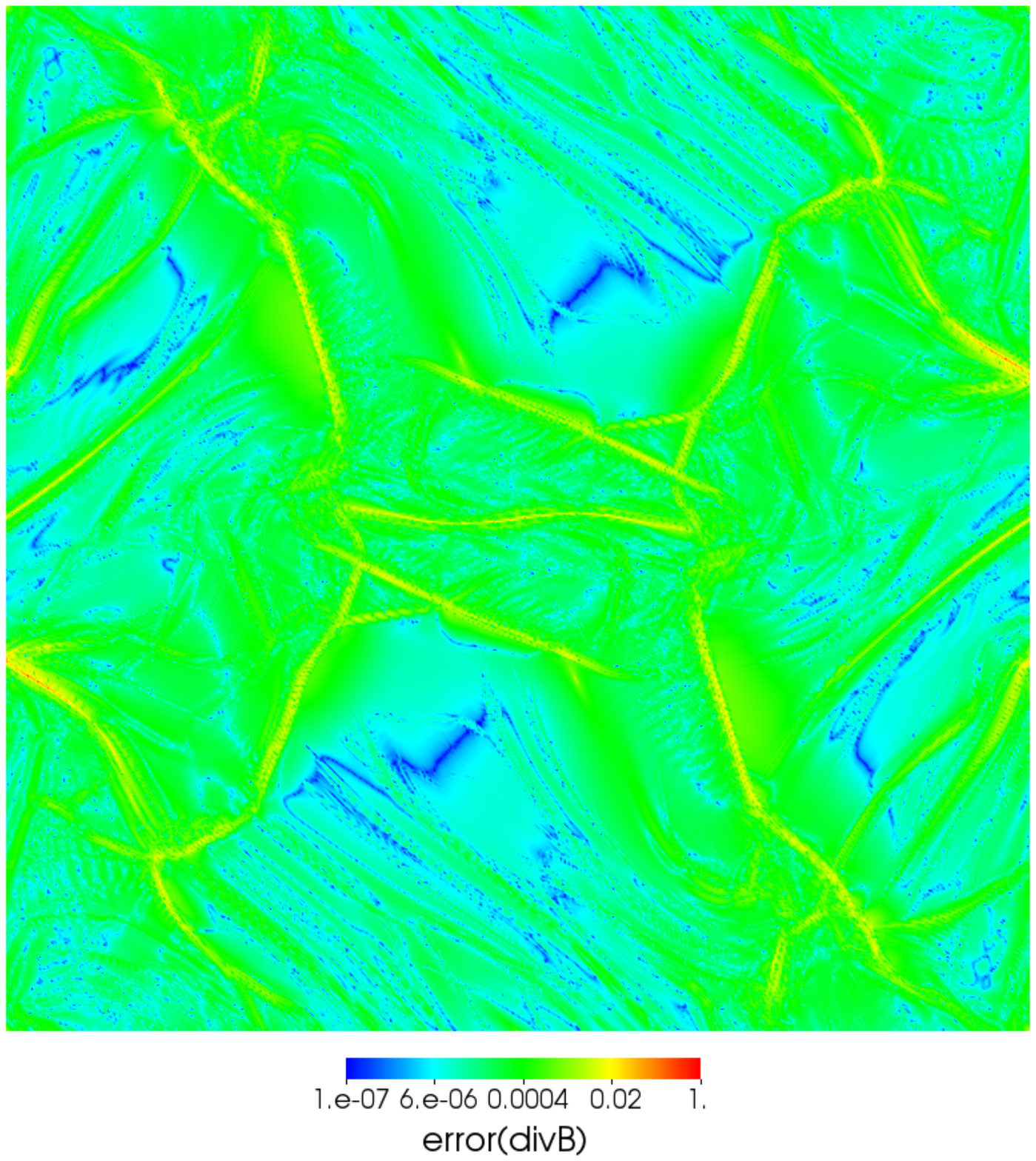} 
\caption{Map of the relative error of $\nabla \cdot {\bf B}$ (see text
  for the definition) in the  Orszag-Tang test at 0.5 time units for our
  simulations at the best resolution.} 
\label{f7}
\end{figure}

\begin{figure}
\centering
\includegraphics[width=8.cm,angle=0]{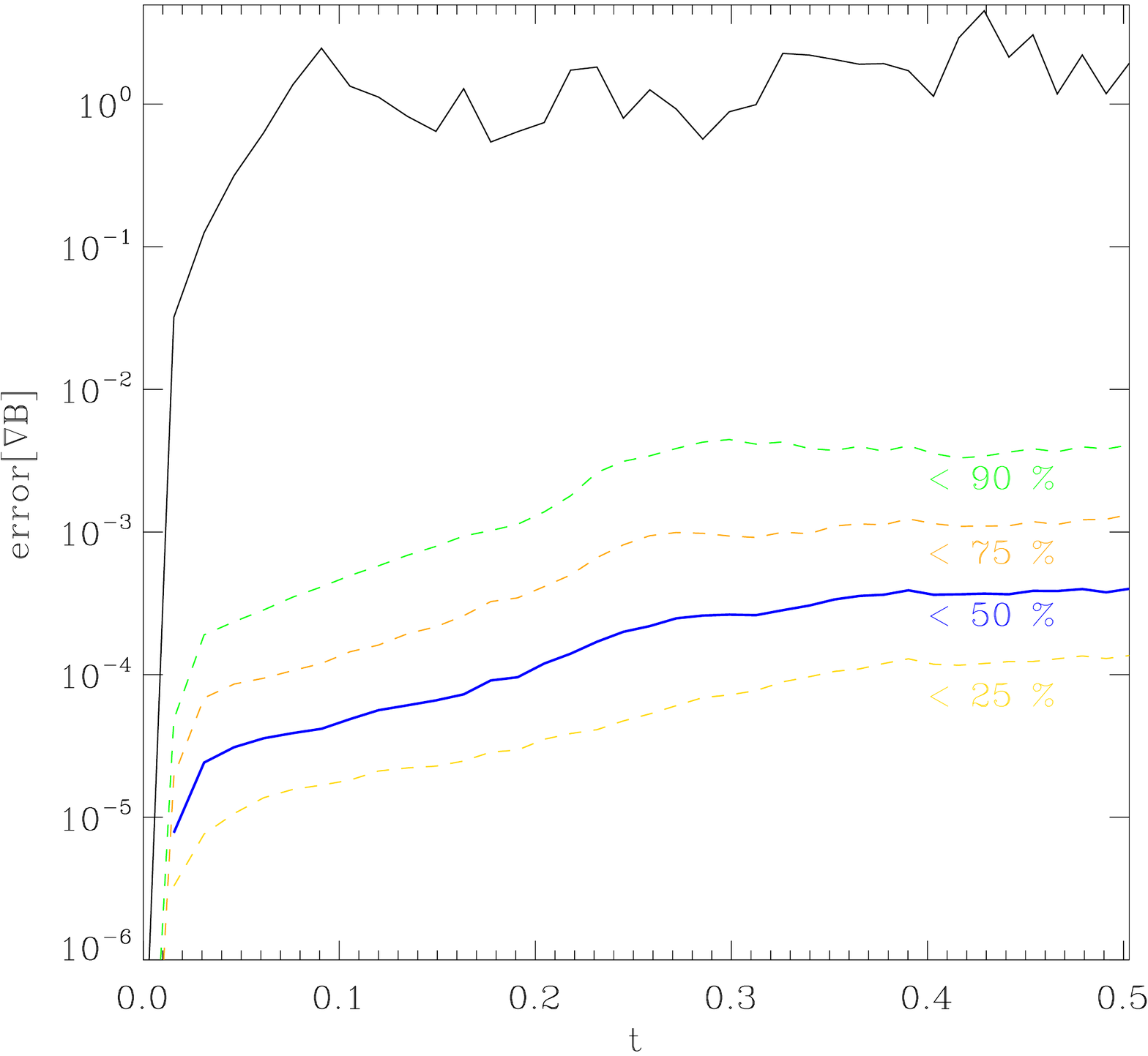} 
\caption{Relative error of $\nabla \cdot {\bf B}$ (see text for the
  definition) in the Orszag-Tang test as a function of time. Yellow,
  blue, orange and green lines indicate the upper bound error for 25,
  50, 75 and 90\% of the cells. The black line represents the maximum
  error of the divergence.}
\label{f8}
\end{figure}

Figure~\ref{f7} displays a map of the relative error of the divergence
of {\bf B} for the best numerical resolution run. As in the fast rotor
test, the largest numerical errors appear associated with flow
discontinuities and do not accumulate in special locations of the
domain. In order to know the time behaviour of the numerical errors,
Fig.~\ref{f8} shows the normalized divergence of the magnetic field
for the AMR run as a function of time. As in the previous test,
yellow, blue, orange and green lines indicate the upper 
bound error for 25, 50, 75 and 90\% of the cells along the
evolution. These errors, within $10^{-4}$ and $10^{-2}$, tend to
stabilize towards the end of the simulation. The black line represents the
maximum error of the divergence (of order 1) along time. Compared with
the divergence error in the rotor test, the median error in this test
is about one order of magnitude larger but still reasonably small
($\sim 0.4\%$). The maximum error compares well with the one obtained by
\cite{GP18} for the same test with an AMR run using divergence
cleaning and is about a factor of ten the one reported by \cite{SD13}
with an SPMHD code using divergence cleaning.

%%%%%%%%%%%%%%%%%%%%%%%%%%%%%%%%%%%%%%%%%%%%%%%%%%%%%%%%%%%
%
%
%%%%%%%%%%%%%%%%%%%%%%%%%%%%%%%%%%%%%%%%%%%%%%%%%%%%%%%%%%%

%%%%%%%%%%%%%%%%%%%%%%%%%%%%%%%%%%%%%%%%%%%%%%%%%%%%%%%%%%%
%

%
%%%%%%%%%%%%%%%%%%%%%%%%%%%%%%%%%%%%%%%%%%%%%%%%%%%%%%%%%%%

%%%%%%%%%%%%%%%%%%%%%%%%%%
\section{A cosmological application: magnetic fields in galaxy clusters}
\label{s:cluster}
%%%%%%%%%%%%%%%%%%%%%%%%%

A straightforward application of a cosmological code would be to
simulate the evolution of a cosmic volume where a large number of  
cosmic structures, spanning a huge range in masses and sizes, form and
evolve. Thus, as an example of application of the new version of the code
MASCLET, we present the results for a simulation of a
computational box representing a moderate volume of the Universe. 

%%%%%%%%%%%%%%%%%%%%%%%%%%%%%%%%%%%%%%%%%%%%%%%%%%%%%%%%%%%%%%%%%%%%%%%%%%%%% 
%
\begin{figure*}
\centering
{\includegraphics[width=8.cm,angle=0]{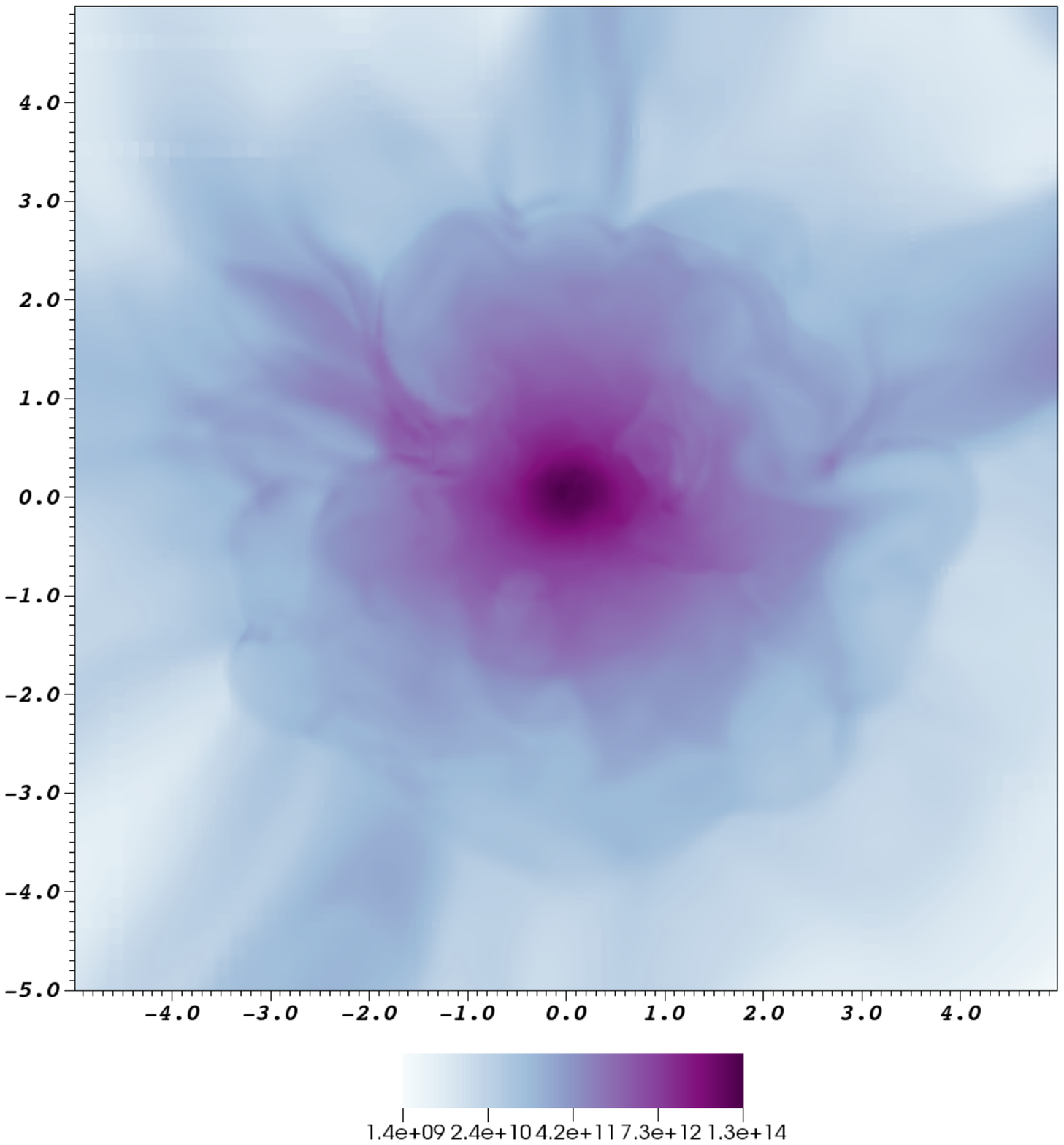}} 
{\includegraphics[width=8.cm,angle=0]{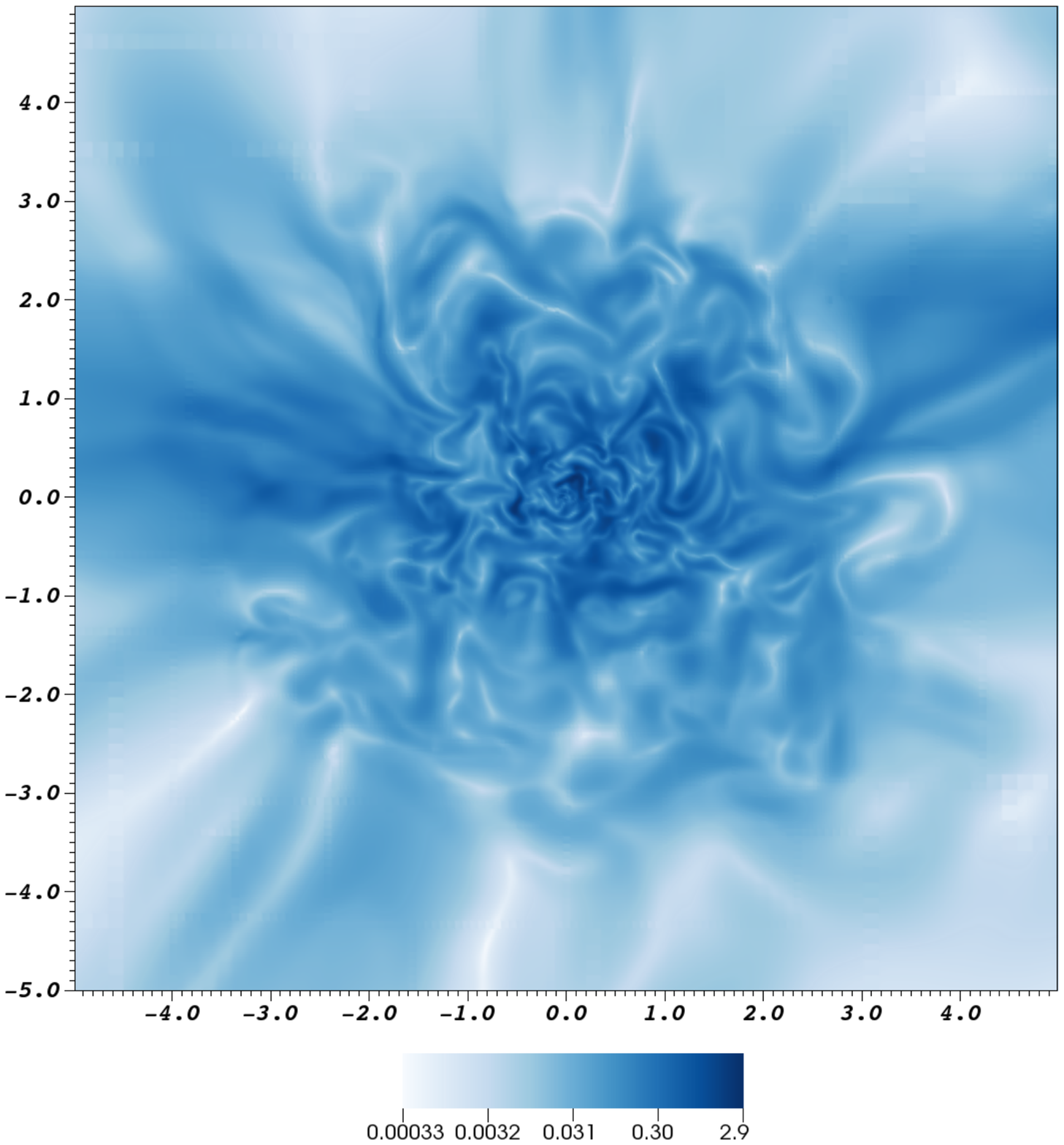}} 
\caption{Slices ($\sim 8$ kpc width) through the cluster center for
  density (left) and B (right) at $z\sim 0$. The density is in units
  of $M_{\odot}$/Mpc$^3$ and the comoving magnetic field intensity,
  $B$, is in $\mu$G. The spatial scales are in comoving Mpc.} 
\label{fig9}
\end{figure*}
%
%%%%%%%%%%%%%%%%%%%%%%%%%%%%%%%%%%%%%%%%%%%%%%%%%%%%%%%%%%%%%%%%%%%%%%%%%%%%%

The simulation assumes a spatially  flat $\Lambda$CDM cosmology, with 
these cosmological  parameters:  $\Omega_{\rm m} = 0.31$,
$\Omega_{\Lambda} = \Lambda/{3H_0^2} = 0.69$, $\Omega_{\rm b} = 0.048$,
$h = H_0/(100$ km s$^{-1}$ Mpc$^{-1}$) $=0.678$,  $n_{\rm s} = 0.96$
and $\sigma_8=0.82$. The simulated region is discretised  with
$128^3$ cubical cells within a cube of comoving  side length $40$
Mpc. Employing a  CDM transfer function from \citet{Eisenstein_1998},
we  set up the initial conditions at  $z = 100$. We applied a
constrained realization in order to generate a rich galaxy cluster in
the center of the box \citep[see][]{Hoffman_1991}. 

From the initial conditions, evolved until present  time using a low
resolution domain, we select regions satisfying some refining criteria
in order to arrange three  refinement levels ($l=1, 2$, and $3$) for
the AMR scheme. In  these initially refined  levels, the  dark matter
(DM)  component is sampled with  DM  particles 8, 64, and 128 times,
respectively, lighter than  those used  to sample regions in the
coarse grid ($l=0$). As the evolution proceeds, the total density,
that is baryonic plus dark matter densities, is used to refine regions
on  the different grids by means of a pseudo-Lagrangian approach in
which a cell is flagged as refinable if its density increases in a
factor of eight.  

In the present simulation we use a maximum of seven refinement levels 
($l = 7$), allowing for a peak physical spatial resolution of $\sim 3$
kpc at $z=0$. Four different particles species are considered for the DM,
corresponding to those particles on the coarse grid and those
within the three first  levels of refinement.  The best  mass
resolution  is  $\sim 2\times  10^6\,  M_\odot$, equivalent to use
$1024^3$ particles in the whole computational domain. 

The simulation is adiabatic, therefore, no cooling or heating
processes are considered. The gas is described as an ideal fluid with
an adiabatic exponent equal to $5/3$. 

We mimic the primordial cosmological magnetic field by seeding a
comoving homogenous magnetic field, ${\bf B}_0$, filling the whole
computational domain and orientated along one of the axis of the
computational box. The chosen value of this seed is 0.1 nG, which 
is just below observational constraints imposed by the analysis of  CMB
data \citep[e.g.][]{Subramanian_2016,Planck_2016}. 
This choice of the initial magnetic field is somehow arbitrary
 and neglects many 
other possible sources of this initial field. Nevertheless, for the purpose of 
this paper and for the sake of comparison with previous works, we consider
this value reasonable enough. Although a different choice could have an impact in the final value of {\bf B} if no evidences of dynamo amplification were found.
Some recent
works have also proved that the initial topology of the magnetic field
has no relevant effects on the formation of cosmic structures such as
galaxies and galaxy clusters
\citep[e.g.][]{Marinacci_2015,Vazza_2017,Vazza_2018,DF_Vazza_2019}. 

In terms of the simplicity of the physics involved (only cosmic
expansion, gravity and magnetohydrodynamics), the mass of the formed
cluster and the effective numerical resolution, our simulation can be 
compared with the one with the highest numerical resolution discussed
in \cite{Vazza_2018}.  
Figure~\ref{fig9} shows two thin slices ($\sim 8$ kpc width) of  the 
gas density (left panel) and the magnetic field strength (right panel)
at $z\sim 0$. The slices cut the cluster centre and have a size of
four virial radius approximately. The density plot exihibits a
prominent core, as expected in an adiabatic 
simulation, corresponding to a galaxy cluster with a mass of $4\times
10^{14}\, M_{\odot}$ and a virial radius $R_{\rm Vir} = 1.96$ Mpc. The
comoving magnetic field intensity, $B$, shows a more complex structure 
as a consequence of the entanglement of the magnetic field lines,
specially within the virial radius. The similarity of this plot
  (characterized by entangled tongues of alternating high and low
  magnetic field) with the corresponding one in \cite{Vazza_2018} (see
  their Figure~2) is remarkable. 

\begin{figure*}
\centering
\includegraphics[width=17.cm,angle=0]{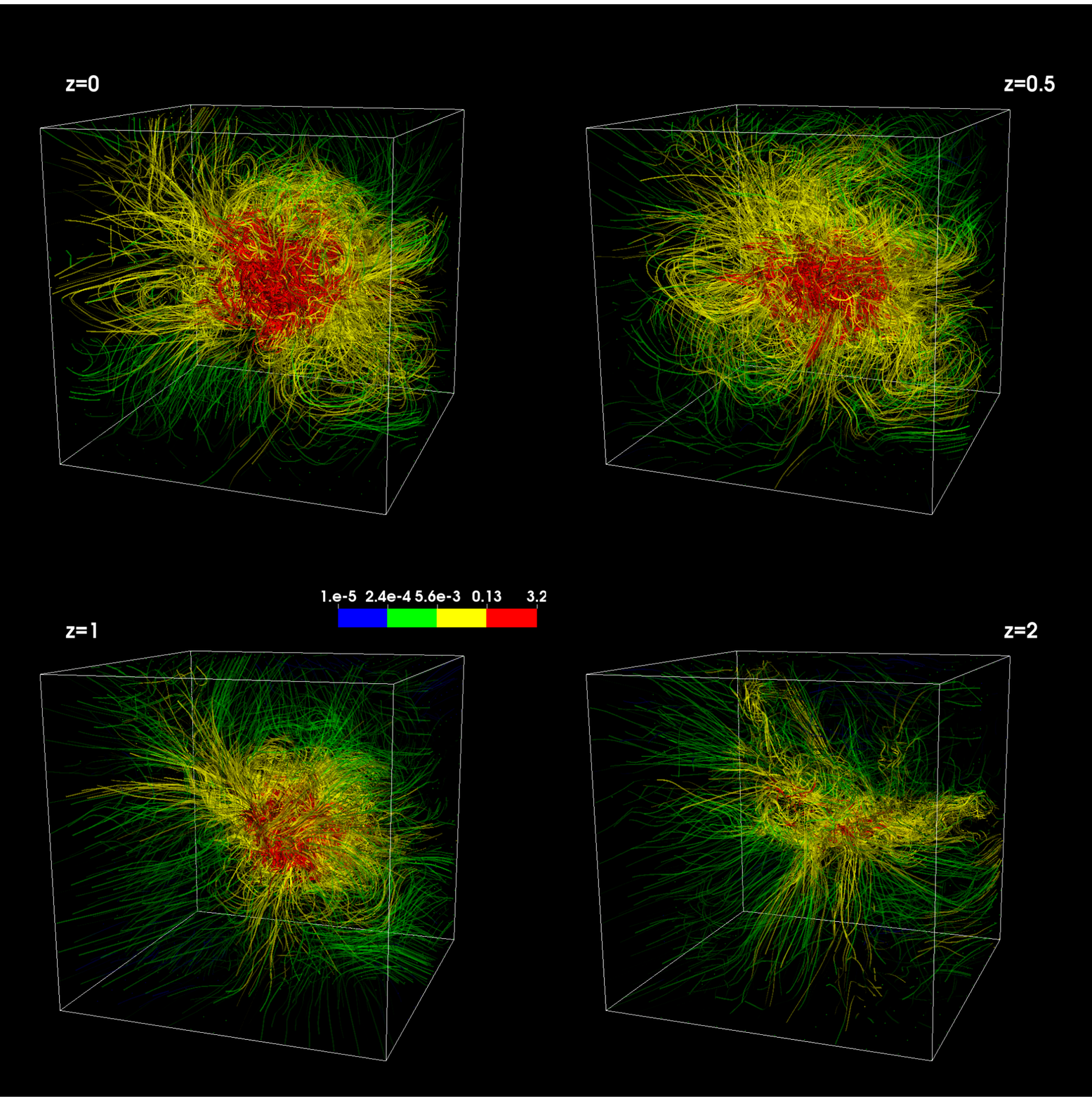} 
\caption{Magnetic field lines distribution at four redshifts, 2, 1,
  0.5, and 0 in a box centred at the cluster with a side length of
  eigth comoving Mpc (four virial radius). The lines are colour-coded
  according with the value of the comoving magnetic field in units of 
  $\mu$G.} 
\label{fig10}
\end{figure*}

In order to visualize the topology and structure of the magnetic field
and its time evolution, we display the Fig.~\ref{fig10}. The four
panels present the magnetic field lines in a cubic region centred at
the cluster position at $z\sim 0$ with a side length of 8 comoving Mpc
($\sim 4 R_{\rm Vir}$). The lines are colour-coded according to the
value of magnetic field intensity in $\mu$G. Four epochs,
corresponding to redshifts $z = $ 2, 1, 0.5, and 0 are displayed.
It is clearly visible how, as the cluster builds up, the magnetic
field is amplified in the region within the virial radius. Based on
the properties of the magnetic field, the cluster can be separated
into two broad regions. Within the virial radius, although the
distribution of magnetic field lines is complex and intricate, the
intensity of ${\bf B}$ is rather homogenous with values of the order of
$\mu$G. By contrast,
at the cluster's outskirts, field lines basically follow the
filaments of gas feeding the cluster, and the value of $B$ in these
regions is remarkably lower compared with values at the inner
parts.   

%%%%%%%%%%%%%%%%%%%%%%%%%%%%%%%%%%%%%%%%%%%%%%%%%%%%%%%%%%%%%%
%
%%%%%%%%%%%%%%%%%%%%%%%%%%%%%%%%%%%%%%%%%%%%%%%%%%%%%%%%%%%%%%%%%

\begin{figure}
\centering
\includegraphics[width=7 cm,angle=0]{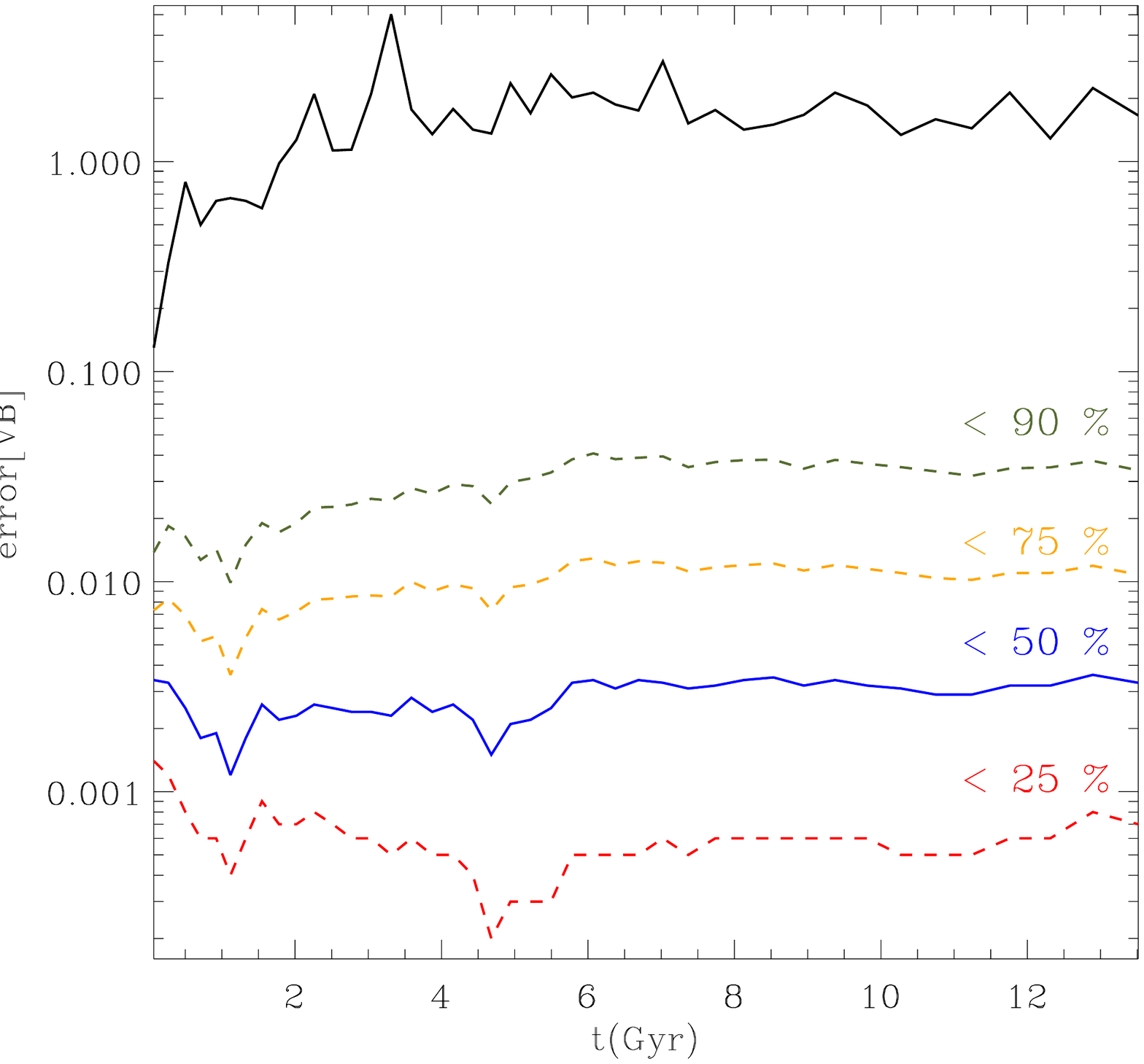} 
\caption{Relative error of  the divergence of the comoving magnetic field
     as a function of cosmic time. Yellow, blue, orange and green
  lines indicate the upper bound error for 25, 50, 75 and 90\% of the
  cells. The black line represents the maximum error of the divergence.}
\label{fig11}
\end{figure}

The radial profile of the magnetic field is presented in
Fig.~\ref{fig11}, where its mass-weighted comoving intensity is
plotted against cluster-centric distance (continuous red line) in units normalized to the
cluster virial radius, $R_{\rm Vir}$. For the sake of comparison, we
also plot  the magnetic field intensity that
would be expected from a pure compression of the magnetic field
lines (discontinuous red line). The values of $B$ in this case can be computed as
$B=B_0(\rho/\bar{\rho})^{2/3}$, where $B_0 = 0.1$ nG is the comoving
magnetic field, and $\rho$ and $\bar{\rho}$ are, respectively, the gas
density and the average gas density in the whole computational box. 

Our results show a $B$ profile which is above the
expected profile for magnetic fields from pure compression of the
gas. The computed magnetic field is in fact 2 to 10 times higher than
the one estimated from pure compression for most of the cluster volume,
$r > 0.1 R_{\rm Vir}$ (at the very inner centre of the cluster, $r<0.1
R_{\rm Vir}$, the profile of $B$ presents a more complex
structure with a peak and a central dip which we associate with the 
particular dynamical state of the simulated cluster). This magnetic
field amplification is in contrast with the results for the
adiabatic simulation displayed by \cite{dubois2008} (see their
Figure~3). In this case, the magnetic field at the inner parts of the
cluster is below the expected values from $B \propto \rho^{2/3}$. The
authors attribute these results to numerical dissipation and invoke
cooling processes -- which resumes gravitational contraction and
shearing motions -- to amplify the magnetic field.

As we already advanced, our results are much more in
agreement with those of \cite{Vazza_2018}. In this work, considering
also an adiabatic simulation with a comparable effective numerical
resolution, the authors manage to amplify the magnetic field to values
of the order of $\mu$G (starting from a seed field of 0.1 nG) and
typically one order of magnitude above the value expected if the field
purely follows the gas ($B \propto \rho^{2/3}$). According to these
authors, at the highest spatial resolution ($\approx 4$ kpc) the
numerical dissipation of their code is small enough to resolve the
small-scale turbulent dynamo process. In this process, the magnetic
field lines are stretched following the fluid particles random motions
until diffusion stops the amplification. It is then critical for the
success of the process that i) numerical viscosity remains small
enough to let the code maintain turbulent motions at the required
spatial scales, and ii) numerical magnetic resistivity remains small
enough to let the turbulent flow stretch the magnetic field lines
efficiently before non-ideal effects kill the process. The
  similarity of the magnetic intensity distribution shown in 
  Fig.~\ref{fig9} with the corresponding one in \cite{Vazza_2018}
  (Fig.~2), might point to the same process as the responsible of the
  additional magnetic field amplification seen in
  Fig.~\ref{fig9}. This should be verified in future works. As a first
  step, we have confirmed that the modulus of the vorticity (not
  shown) and the magnetic field intensity are closely correlated,
  pointing to turbulence as the origin of the added field
  amplification. 

To assess the performance of our code in keeping magnetic
field divergence errors under control, and as in previous sections, we
study the normalized divergence of the comoving magnetic
field, ${\bf B}$. In this particular test, we modify the definition
of this quantity to reduce the weight of spurious values associated to
cells where the magnetic field intensity is very low. Following
\cite{Tricco2012}, we change $|{\bf B}_i|$ in the denominator of the
normalized divergence by $|{\bf B}_i| + 0.01 |{\bf B}_{\rm max}|$,
where $|{\bf B}_{\rm max}|$ is the maximum of $|{\bf B}_i|$ across the
numerical grid.

Figure~\ref{fig11} displays the median of the relative error (blue 
thick line), and the relative error of the 25\% (yellow line), 75\%
(orange line), and 90\% (green line) percentiles, respectively. The
black line shows the maximum relative error of $\nabla \cdot {\bf B}$ 
according to the previous definition. Finally, let us note that with
90\% of the numerical cells with a divergence relative error well
below $3\times 10^{-2}$ in $\nabla \cdot {\bf B}$, the spurious
magnetic energy change related to the divergence error must be of the
order of $\sim 10^{-3}$ or less, i.e., much smaller than the increase
of magnetic energy associated to the added magnetic field
amplification.

%%%%%%%%%%%%%%%%%%%%%%%%%%%%%%%%%%%%%%%%%%%%%%%%%%%%%%%
%
%%%%%%%%%%%%%%%%%%%%%%%%%%%%%%%%%%%%%%%%%%%%%%%%%%%%%%%%

\begin{figure}
\centering
\includegraphics[width=8 cm,angle=0]{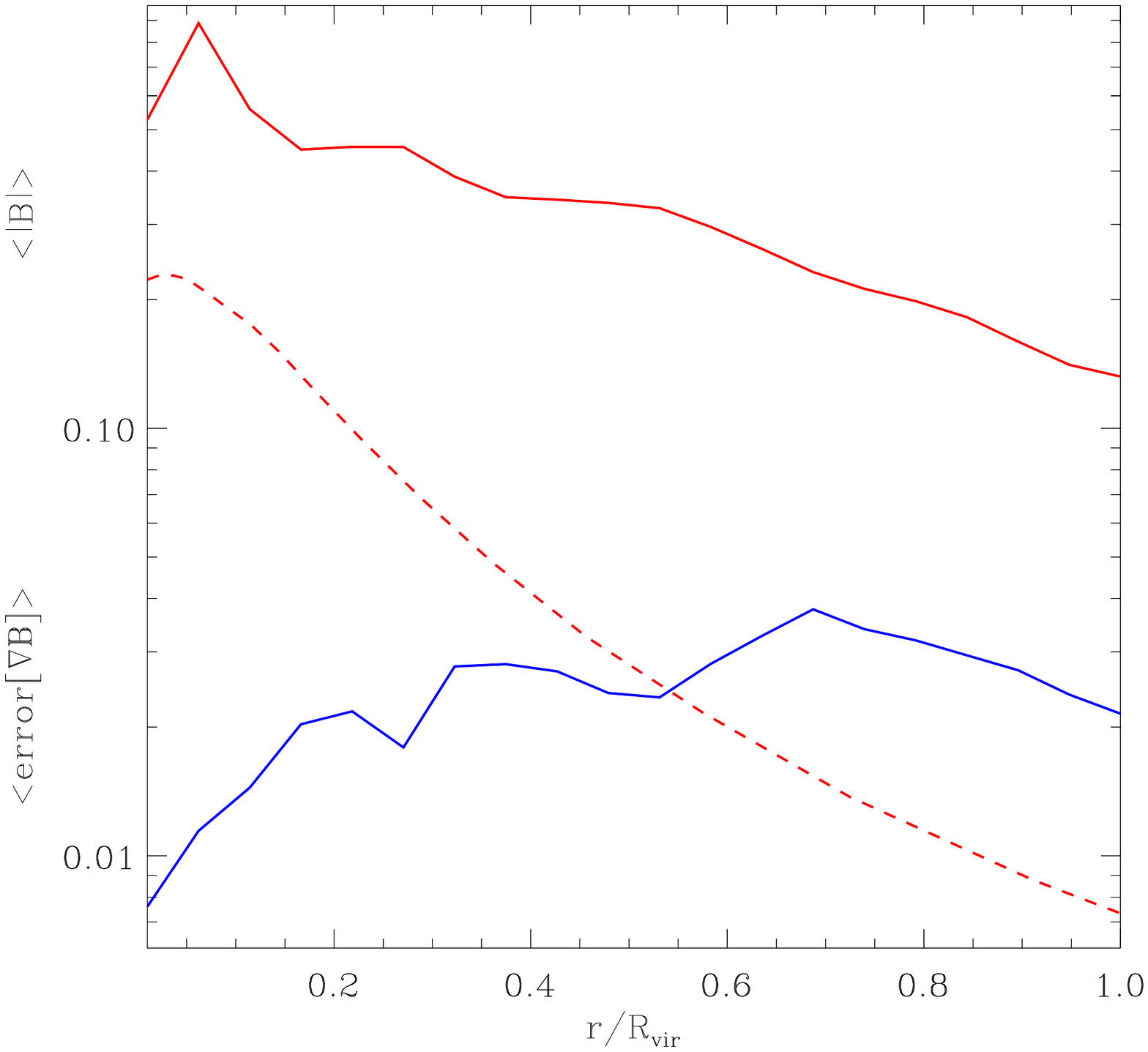} 
\caption{Radial profile of the comoving magnetic field intensity (in $\mu G$)  as a
  function of the cluster-centric distance in units of virial
  radius (red lines).  The continuous line corresponds to the profile obtained
  from the simulation and the red dashed line represents the value of
  B expected from an amplification of the magnetic field from a pure
  compression defined as $B=B_0(\rho/\bar{\rho})^{2/3}$. The solid blue line represents the radial profile of the average relative error of the divergence of {\bf B} in the same radial units}. 
\label{fig12}
\end{figure}

The most relevant structure formed in this computational box is
the galaxy cluster located at the box centre. The radial profile of the average value  of the error of the divergence of
{\bf B} as a function of the cluster-centric distance is displayed as the blue line in Fig.~\ref{fig12}. The
average values are computed as volume weighted means in spherical
shells. 
This average radial error can be compared with the average radial value of $|{\bf B}|$ from the simulation together with the expected value of the magnetic field intensity from a pure compression (red lines in Fig.~\ref{fig12}).
The
relative error exhibits an acceptable behaviour with values below 4\%.

%%%%%%%%%%%%%%%%%%%%%%%%%%%%%%%%%%%%%%%%%%%%%%%%%%%%%%%
%

\begin{figure}
\centering
\includegraphics[width=9. cm,angle=0]{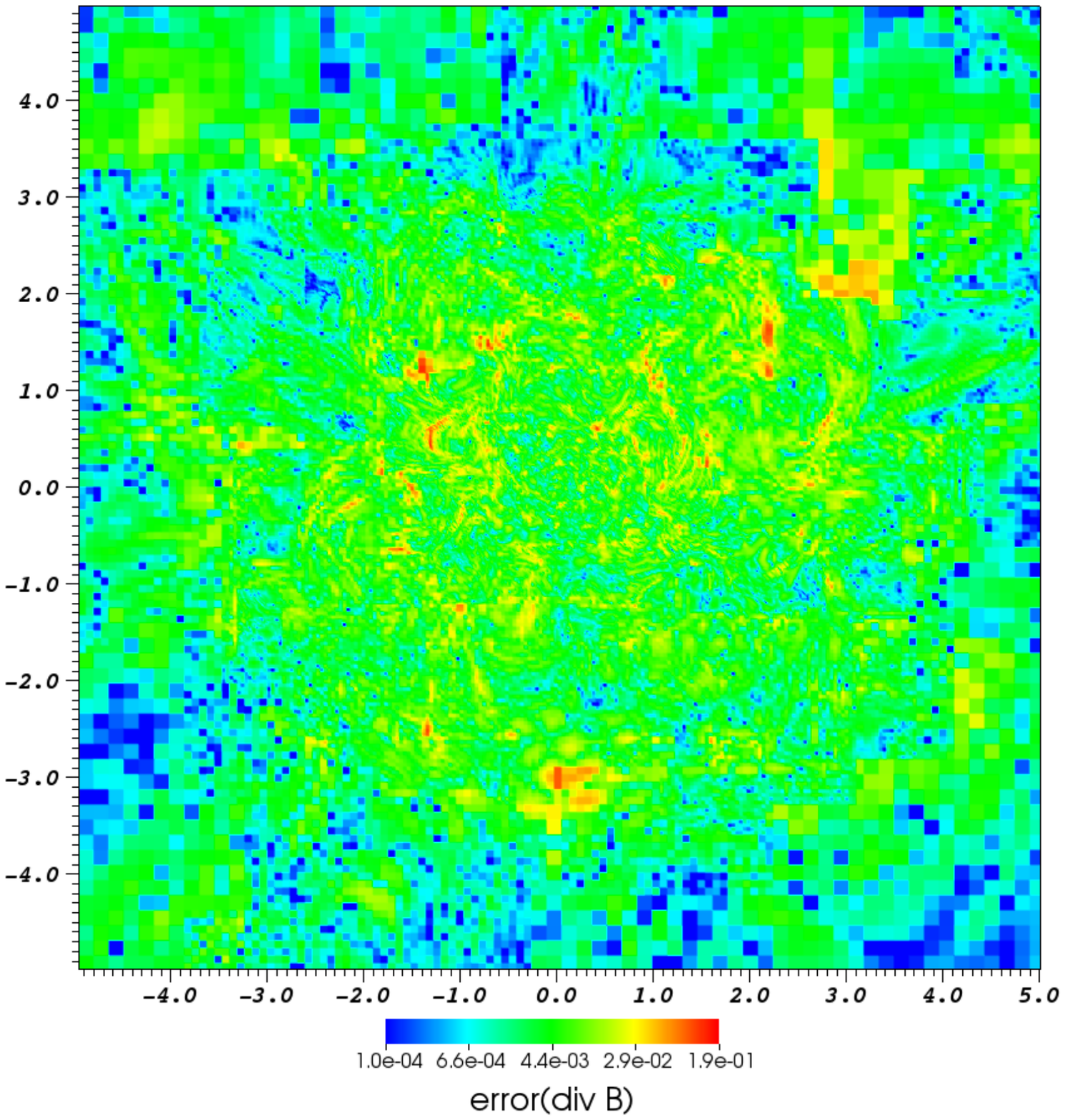} 
\caption{Slice ($\sim 8$ kpc width) through the cluster center for the
  relative error of the divergence of {\bf B}.} 
\label{fig13}
\end{figure}

To spot possible large errors that could be concealed in the radial
average values of the normalized divergence, we present in 
Figure~\ref{fig13} a map of the error of the divergence of $B$ in the
same thin slice through the cluster centre displayed in
Fig.~\ref{fig9}. As seen in the figure, the point values of the
normalized divergence in the central region of the cluster are smaller
than 20\% with a mean value $~0.5\%$, in agreement with the median
error shown in Fig.~\ref{fig11}. 

%%%%%%%%%%%%%%%%%%%%%%%%%%%%%%%%%%%%%%%%%%%%%%%%%%%%%%%
%%%%%%%%%%%%%%%%%%%%%%%%%%%%%%%%%%%%%%%%%%%%%%%%%%%%%%%%

%%%%%%%%%%%%%%%%%%%%%%%
\section{Conclusions}
\label{s:conclusion}
%%%%%%%%%%%%%%%%%%%%%%%%%

The role of magnetic field in cosmological scenarios has traditionally
been a matter of debate. Most of the times, they have been pushed into
the background as minor actors in the formation and evolution of
cosmic structures, whose part was completely subdominant. However,
this picture is rapidly changing in recent years. In the epoch of the
so called precision cosmology, where theory and data are widening our
knowledge on the formation of galaxy and galaxy clusters, the
magnetic fields can play a relevant role, both in the theoretical and
observational planes.   

In this line, we presented a new version of an already well-tested and
used cosmological code, that includes a proper description of the MHD
processes. Several commonly used tests, whose solutions are perfectly
known, are shown. In all of them, the performance of our code is more
than satisfactory.  

As a final application with cosmological interest, we followed the
evolution of gas and dark matter components in a cosmic volume, where
a uniform magnetic field was introduced at a very early stage of the
evolution ($z\sim100$). In this simulation, the magnetic field was
amplified from its initial uniform value of ${B_0} \sim 0.1$ nG
until values of the order of ${B} \sim 1 \, \mu$G in the most
massive galaxy clusters. Our simulation shows how the magnetic field
is channeled along the filament of gas and it is amplified at the
structures formed at the intersections of such filaments. The topology
of the magnetic field lines clearly reveals how the field lines tangle
up in the galaxy clusters. The comparison between the values of
$B$ obtained in the simulation and those expected from pure
compression ($B \propto \rho^{2/3}$) reveals that there is an
additional process operating at the core of the cluster responsible
for the extra amplification of the magnetic field. To elucidate the
nature of this process, essential to understand the origin of magnetic
fields in clusters of galaxies, will be the subject of future research.

It is nowadays clear that cosmic magnetic fields, although dynamically
negligible, play a crucial role in shaping the physical properties of
the intergalactic medium. From an observational point of view, current
radio observations, such as those with JVLA or LOFAR, provide a unique
tool to reveal the current distribution of magnetic fields
\citep[e.g.][and references therein]{Donnert_2018}. In this sense,
current and next generation of radio facilities, like the SKA and its
precursors, will achieve, mainly by means of  the Faraday tomography
technique, an unprecedented detail in the description of the evolution
and distribution of cosmic magnetic fields. From a theoretical point
of view, while in the last years cosmological simulations have been
significantly improved in terms of the development of different
feedback models to explain galaxy formation, a proper numerical
understanding of cosmic magnetic fields and gas turbulence phenomena
remains as a challenge. Therefore,   in order to  interpret future
radio observations and to shorten the existing distance between the
observational and the theoretical planes, an accurate and `realistic'
description of cosmic magnetic fields in full cosmological simulations
seems to be imperative.

\section*{ACKNOWLEDGEMENTS}
{This work has been supported by the Spanish Ministerio de Ciencia,
Innovaci\'on y Universidades (MICINN, grant AYA2016-77237-C3-3-P) 
and by the Generalitat Valenciana (grant PROMETEO/2019/071). JMM
also acknowledges partial support from MICINN (grant
PGC2018-095984-B-I00). 
The authors are indebted to D. Ryu and T.W. Jones for providing them
with the analytical solutions for the one-dimensional tests shown in
the paper and ackownledge fruitful discussions with
P. Cerd\'a-Dur\'an. We thank the anonymous referee for his/her
constructive criticism. Simulations have been carried out using the
supercomputer Llu\'is Vives at the Servei d'Inform\`atica of the
Universitat de Val\`encia.} 

\bibliographystyle{mnbst}
\bibliography{mnras_masclet-b} 

%\appendix

\label{lastpage}

\end{document}